\documentclass[a4paper,11pt]{article}
\usepackage[T1]{fontenc}
\usepackage{lmodern}
\usepackage{mathtools, siunitx, physics, array, bm, graphicx, bbold}
\usepackage{amsmath}
\usepackage{dsfont}
\usepackage{booktabs}
\usepackage{multirow}
\usepackage{tikz}
\usepackage{wrapfig}
\usepackage[compat=1.1.0]{tikz-feynman}
\usepackage{subcaption, caption}
\usepackage{adjustbox}
\usepackage{enumitem, comment}
\usepackage{slashed}
\usepackage{float}
\usepackage{pdfpages}
\usepackage{jheppub}
\usepackage{academicons}
\usepackage{mathrsfs}
\usepackage{orcidlink}

\tikzfeynmanset{
every crossed dot@@/.style={
    /tikz/fill=none,
    /tikz/draw=black,
    /tikz/shape=crossed circle,
    /tikz/minimum size=0.25cm,
    /tikz/inner sep=0pt,
},
wblob/.style={
    /tikz/fill=none,
    /tikz/draw=red,
    /tikz/shape=crossed circle,
    /tikz/minimum size=0.25cm,
    /tikz/inner sep=0pt,
},
gblob/.style={
    /tikz/fill=none,
    /tikz/draw=blue,
    /tikz/shape=crossed circle,
    /tikz/minimum size=0.25cm,
    /tikz/inner sep=0pt,
},
wgblob/.style={
    /tikz/fill=none,
    /tikz/draw=green!50!black,
    /tikz/shape=crossed circle,
    /tikz/minimum size=0.3cm,
    /tikz/inner sep=0pt,
},
}

\usetikzlibrary{positioning,arrows.meta}
\tikzfeynmanset{ with arrow/.style = {
   decoration={
     markings,
     mark=at position 0.5
          with {\arrow[xshift=1mm]{Triangle[black,width=0.8mm,length=1.2mm]}}
     },
   postaction=decorate}
}

\title{Axion EFT in the BMHV Scheme: Flavor Currents, Evanescent Operators and Ward Identities}

\author{Deepanshu Bisht\orcidlink{0009-0009-7047-773X},}
\author{Sabyasachi Chakraborty\orcidlink{0000-0001-5356-7607},}
\author{and Atanu Samanta\orcidlink{0009-0006-2782-2911}.}

\affiliation{Department of Physics, Indian Institute of Technology Kanpur, Kanpur-208016, India}

\emailAdd{dbisht22@iitk.ac.in}
\emailAdd{sabyac@iitk.ac.in}
\emailAdd{asamanta23@iitk.ac.in}

\abstract{We present a systematic analysis of axion effective field theory within the Breitenlohner–Maison–’t Hooft–Veltman (BMHV) scheme, focusing on the renormalization of fermionic dimension-five operators and the associated chiral flavor currents. In this framework, the non-anticommuting nature of $\gamma_5$ in $d \neq 4$ dimensions leads to violations of naive Ward identities through the emergence of evanescent operators. We derive the bare and renormalized Ward identities for chiral currents, explicitly identifying the equation-of-motion and evanescent operator contributions. Using diagrammatic calculations, we verify the validity of these identities up to two-loop order $\mathcal{O}(\alpha_s^2)$, including both pole and finite terms. We demonstrate how evanescent operators mix into physical operators and determine the finite renormalization required to restore four-dimensional Ward identities, recovering the expected structure of axial current renormalization and the anomaly. Our results provide a consistent and transparent framework for multi-loop computations in axion EFT and highlight the essential role of evanescent operators in maintaining scheme consistency.}

\begin{document}
\maketitle
\flushbottom

\section{Introduction}
It is a well-established principle that any scheme dependence arising in quantum field theory calculations must cancel in physical observables. This includes, for example, the dependence on the renormalization-scheme, associated with the choice of counterterms used to subtract ultraviolet divergences, as well as the dependence of the gauge, which must disappear from all measurable quantities. In this sense, regularization-scheme dependence is no exception. Among the available methods, dimensional regularization is particularly popular due to its powerful theoretical properties and its practicality in multi-loop computations. However, dimensional regularization itself admits several a priori non-equivalent prescriptions. With the steadily increasing precision of modern high-energy experiments, it has become essential to scrutinize the robustness of theoretical predictions obtained within specific dimensional-regularization schemes. In particular, one must verify that the physical observables and amplitudes computed in different schemes agree to any fixed order in perturbation theory. 

The most commonly used regularization prescription is {\it Naive Dimensional Regularization} (NDR)~\cite{Chanowitz:1979zu}, in which both the Dirac algebra and the spacetime metric are continued to $d = 4  - 2\epsilon$ dimensions. Although this approach is computationally convenient, the extension of the chiral matrix $\gamma_5$ to $d \neq 4$ dimensions is inherently ambiguous~\cite{Martin:1999cc} (see Appendix ~\ref{app:NDR} for more details). In contrast, the {\it Breitenlohner–Maison–’t Hooft–Veltman} (BMHV) scheme~\cite{tHooft:1972tcz,Breitenlohner:1975hg, Breitenlohner:1976te, Breitenlohner:1977hr} provides a mathematically consistent framework by explicitly splitting the Dirac algebra into a four-dimensional subspace (represented by {\it tilde}) and an evanescent $\epsilon$-dimensional component such as
\begin{equation}
     \gamma^\mu = \widetilde{\gamma}^\mu +  \widehat{\gamma}^\mu \;, \quad g^{\mu\nu} = \widetilde{g}^{\mu\nu} + \widehat{g}^{\mu\nu} \;, \quad p^\mu = \widetilde{p}^\mu + \widehat{p}^\mu\;,
\end{equation}
see Appendix~\ref{app:HValgebra} for more details on the algebra and relevant identities of the BMHV scheme. This construction ensures consistency to all orders in perturbation theory. Nevertheless, the BMHV scheme introduces additional technical complications: in particular, the explicit separation of dimensions leads to the breaking of Ward/Slavnov–Taylor identities. These violations must be compensated for by the introduction of suitable finite counterterms in order to restore the symmetries of the theory and obtain physically meaningful results. 
    
In recent years, conceptual ingredients necessary for the consistent application of the BMHV scheme have received a great deal of attention. For example, a systematic analysis of the BMHV scheme in the context of renormalizable chiral gauge theories was presented recently in~\cite{Belusca-Maito:2023wah}. The main motivation for that work was the conceptual importance of employing a consistent and robust dimensional regularization framework. Within this approach, the Slavnov Taylor identities are imposed to determine the finite counterterms required to restore gauge symmetry order by order in perturbation theory. In a related study, Ref.\cite{Cornella:2022hkc} derived the one-loop finite counterterms of the Standard Model in the BMHV scheme using the background field method. In addition, a systematic discussion of evanescent operators in effective field theory (EFT) was provided in \cite{Fuentes-Martin:2022vvu}. Together, these works establish the conceptual ingredients necessary for applying the BMHV scheme in modern applications. Although the emphasis is primarily on formal aspects, further developments are required to facilitate their use in concrete phenomenological calculations.

In the recent past, given the widespread use of effective field theories in phenomenological studies involving Standard Model (SM) or Beyond-the-standard model (BSM) frameworks, it is important to employ a robust and self-consistent regularization scheme, such as the BMHV, in practical EFT calculations. This also provides a natural framework for a systematic analysis of the scheme dependence. Indeed, much of the early literature on SM processes and weak decays relevant for low-energy collider phenomenology pursued similar objectives~\cite{Buras:1989xd, Buras:1992zv}. These studies focused on \textit{effective} dimension-six operators and their loop corrections. In this context, evanescent operators were first identified and incorporated into practical EFT computations. Their appearance was associated with the breakdown of {\it Fierz} identities, which are no longer valid in dimensionally regularized theories away from four dimensions. The inclusion of evanescent operators was shown, for example, to affect the anomalous dimensions of physical operators starting at the two-loop level. The general relations between the scheme-dependent quantities were then derived and verified diagrammatically, ultimately obtaining reliable, scheme-independent physical results~\cite{Ciuchini:1993ks, Ciuchini:1993vr}. More recent work has emphasized the systematic identification and treatment of {\it Fierz-related} evanescent operators in effective theories such as SMEFT and LEFT~\cite{Fuentes-Martin:2022vvu, Naterop:2023dek, Naterop:2025lzc}, often employing functional techniques~\cite{Fuentes-Martin:2023ljp}. On the other hand, in \textit{renormalizable} settings where dimensional regularization was used in deep-inelastic scattering, it was recognized that a robust treatment of $\gamma_5$ is needed to renormalize fermion currents. The foundational work by Collins~\cite{Collins:1984xc} provided a transparent understanding by demonstrating that BMHV breaks current Ward identities through explicit tree-level evanescent operators living in the $\epsilon$-dimensional subspace. By computing one-loop $\mathcal{O}(\alpha_s)$ insertions of the evanescent operators into matrix elements with external fermions, an extra \textit{finite} renormalization of the $\overline{\text{MS}}$-renormalized currents is obtained and the four-dimensional renormalized Ward identities are restored in BMHV. With one-loop finite renormalized current, \cite{Bos:1992nd} soon explicitly verified the Ward identity (WI) at two-loop $\mathcal{O}(\alpha_s^2)$ using matrix elements with gluon final states. In a classic study~\cite{Larin:1993tq}, Larin computed finite renormalization of the currents up to $\mathcal{O}(\alpha_s^3)$ with gluon final states in a $\gamma_5$ prescription by imposing finite renormalized currents to satisfy renormalized four-dimensional Ward identities. This method has been developed up to $\mathcal{O}(\alpha_s^5)$ \cite{Chen:2022lun}, although thus far they have been applied primarily to Standard Model perturbative calculations or lattice QCD studies~\cite{Martinelli:1994ty, Bhattacharya:2015rsa}.

A possible application of the BMHV scheme arises in the study of axions and axion-like particles (ALP), which also have attracted considerable interest due to their potential to resolve the strong-$CP$ problem~\cite{tHooft:1976rip}, their role as viable candidates for cold dark matter~\cite{Preskill:1982cy,Dine:1982ah,Abbott:1982af}, their potential relevance to matter-antimatter asymmetry~\cite{Co:2019wyp,Chakraborty:2021fkp}, and implications for the hierarchy problem~\cite{Graham:2015cka,Hook:2016mqo,Trifinopoulos:2022tfx}. Moreover, recent searches at flavor factories have renewed interest in moderately heavy axions with masses $m_a \sim \mathcal{O}(\text{GeV})$~\cite{Bauer:2017ris,Bauer:2021mvw,Chakraborty:2021wda,Bertholet:2021hjl,Bisht:2024hbs}. Although existing studies have largely relied on the NDR scheme to determine renormalization group evolution (RGEs) and Wilson coefficients through matching at the electroweak scale, in this work, we employ for the first time in the axion context, the BMHV scheme. We make a systematic effort to incorporate relevant evanescent operators in axion EFT calculations. Starting from the effective Lagrangian defined at the Peccei-Quinn symmetry breaking scale $f_a$, we compute the renormalization-group evolution up to two-loop order $\mathcal{O}(\alpha_s^2)$. Apart from theoretical consistencies, these two-loop effects are of potential phenomenological significance when the axion couples to fermions with $\mathcal{O}(1)$ couplings, as noted, for example, in~\cite{Bauer:2020jbp}. 
 
We begin by demonstrating how fermion currents arise naturally in such effective theories as components of dimension-five operators. Consequently, familiar effects such as triangle-anomaly couplings emerge naturally in ALP effective theories, further underscoring the importance of employing robust regularization schemes and of carefully analyzing scheme dependence. Following the methodology of~\cite{Collins:1984xc, Bos:1992nd}, we adopt the BMHV scheme to calculate the matrix elements and the anomalous dimensions diagrammatically in axion EFT. Given the tractable number of relevant Feynman diagrams at this order, we present many aspects of the calculations that can be used to verify our results. Our goal is to present a transparent and coherent framework and extract robust results. Throughout, we rely on principles such as symmetries, Ward identities, and hermiticity, while carefully elaborating theory-specific features, conventional choices, and literature results. We also anticipate extending this framework to include electroweak gauge-boson corrections in future work, an aspect that remains largely unexplored in the context of flavor-current renormalization, with or without BSM.

The organization of this paper is as follows. In Sec.~\ref{sec:flavALPEFT}, we provide a brief overview of how flavor currents and their renormalization arise in the context of the effective field theory of ALP. In Sec.~\ref{sec:WIs}, we present a systematic analysis of global flavor symmetries, the associated currents, and their bare and renormalized Ward identities containing evanescent operators in the BMHV scheme. Subsequently, in Sec.~\ref{sec:axialresults} we present our results of one- and two-loop calculations done in the BMHV scheme, explicitly verifying the bare Ward identities up to $\mathcal{O}(\alpha_s^2)$. In Sec.~\ref{sec:ADM} we derive renormalized four-dimensional Ward identities with physical current and anomaly operators from our two-loop results. The relevant anomalous dimensions of the physical operators satisfying relations from the Ward identities are also obtained. Finally, we conclude in Sec.~\ref{sec:conclusion}.

\section{Axion Effective Theory and Flavor Symmetry}\label{sec:flavALPEFT}
We begin with the effective Lagrangian of axions defined at the UV scale 
\begin{equation}\label{eq:UVlag}
\mathcal{L} = \mathcal{L}_{\text{\tiny SM}} + \frac{1}{2} \left(\partial_\mu a\right)^2 -\frac{m_a^2}{2} a^2 + \sum_{n} \mathcal{C}_n \mathcal{O}_n\;,
\end{equation}
where the leading-order, mass-dimension five operators are defined as
\begin{equation}
\begin{aligned}
    \mathcal{O}_G  \equiv \frac{a}{f_a}\; G_{\mu\nu}^b \widetilde{G}^{\mu\nu b} \;, \quad
    \mathcal{O}^{(\psi)}_{ ij} \equiv \frac{\partial_\mu a}{f_a} \bar{\Psi}_i \gamma^\mu  \Psi_j\;.
\end{aligned}
\end{equation}
Here, $\Psi$ denotes different SM chiral fermion multiplets, $\Psi = (Q_L, u_R, d_R, L_L, e_R)$ and $i,j = \{1,2,3\}$ are generation indices~\footnote{For this work, we do not consider any Wess-Zumino-Witten interactions of axions~\cite{Chakraborty:2024tyx,Bai:2024lpq}}. The Wilson coefficients $\mathcal{C}_{(\Psi)}$ of the fermionic operators are hermitian $3\times 3$ matrices in the generation space. Their real and imaginary parts give rise to CP-even and odd axion couplings, respectively~\cite{Balkin:2025rqe}. The Lagrangian Eq.~\eqref{eq:UVlag} written in the flavor basis is manifestly invariant under the SM gauge group $SU(3)_c \times SU(2)_L \times U(1)_Y$.

To obtain meaningful low-energy predictions, the effective operators must be renormalized, which in turn determines the renormalization-group evolution of the Wilson coefficients. For this purpose, it is instructive to write the fermionic effective operators as a product of a BSM current and a SM flavor current,~\cite{Arteaga:2018cmw}
\begin{eqnarray}
    \mathcal{O}^{(\psi )}_{ij} &&\sim J_\mu^{(\text{BSM})}\; J^{\mu}_{ij} \;; \nonumber \\
     J_\mu^{(\text{BSM})} = \frac{\partial_\mu a}{f_a} \;, &&\quad J^\mu_{ij} =  \bar{\Psi}_i \gamma^\mu  \Psi_j\;.
\end{eqnarray}
Both the BSM and the fermionic SM currents are singlets under the SM gauge group. Similarly, the gluonic operator $\mathcal{O}_G$ can also be expressed as a total derivative of the corresponding {\it Chern-Simons} current, with its Wilson coefficient $\mathcal{C}_G\sim \alpha_s/8\pi$, fixed by matching at the UV scale. Restricting to these five-dimensional leading operators, the BSM current $J_\mu^{\text{(BSM)}}$ does not contribute to operator renormalization. This follows because the ALP is a Standard Model singlet and interacts with SM fields only through higher mass-dimension effective operators suppressed by powers of $\Lambda_{\text{UV}}$. Consequently, ALP interactions do not induce operator renormalization at a fixed mass dimension. In particular, the ALP field renormalization can be ignored. Renormalization at mass-dimension five is therefore entirely governed by the renormalization of flavor currents $J^{ij}_\mu$ due to SM interactions. 

Let us see how these currents arise from a symmetry when Yukawa interactions are switched off. In this limit, the fermions are massless, and there is no distinction between flavor and mass basis. The renormalizable part of the Lagrangian in~\eqref{eq:UVlag} then exhibits a large global flavor symmetry $U(3)^5$ corresponding to chiral rotations of the five chiral multiplets $\Psi$ in their \textit{generation space}. It is convenient to work in the `mass' basis with Dirac-type fermions. We then express the global $U(3)$ symmetry transformations acting on a given Dirac fermion species $\psi$ as:  
\begin{equation}\label{eq:chiraltransformation}
    \psi_i \to \psi'_i =  \exp\bigg[i (\alpha^a_L P_L + \alpha^a_R P_R) t^a_{ij} \bigg]\psi_j \;,
\end{equation} 
where $P_{L,R} = (1 \mp \gamma_5)/2$ are the chiral projection operators and $t^a$'s are the generators of the $U(3)$ transformations having generation indices $i, j$. These are Hermitian but not necessarily traceless, and $a = \{1, \hdots, 9 \}$. The sum over repeated indices is implied. These transformations act on up- and down-type quarks and leptons. The Noether currents corresponding to this internal symmetry are
\begin{equation}\label{eq:currentsLR}
    (j^a_L)^\mu = \bar{\psi}_i P_R \gamma^\mu P_L t^a_{ij} \psi_j \;, \quad  (j^a_R)^\mu = \bar{\psi}_i P_L \gamma^\mu P_R t^a_{ij} \psi_j \;,
\end{equation}
where the SM gauge group indices are implicit. These are nothing but the SM flavor currents to which the ALP couples, once we make the following identifications
\begin{equation}
    \mathcal{C}_{Q_L} \to \text{diag}(c^a_{u L}, c^a_{d L})t^a \;, \quad \mathcal{C}_{u_R} \to c^a_{u R} t^a \;,\quad \mathcal{C}_{d_R} \to c^a_{d R} t^a \;,
\end{equation}
and similarly for the leptons. 
Since the generators $t^a$ constitute the basis for $3\times 3$ Hermitian matrices, any $\mathcal{C}^{(\psi)}$ can be formed using their appropriate linear combinations. Note that naively it seems that in the mass basis, there are seven $U(3)$ transformations: six for left- and right-handed up and down-type quarks and charged leptons, and one for left-handed neutrinos. However, constraints imposed by $SU(2)_L$  gauge invariance reduce the number of independent transformations to five. This is also reflected in the mass-basis Wilson coefficients of ALP interactions with left-handed SM fermions~\cite{Bisht:2024hbs}. 
In conclusion, the ALP is coupled to the $U(3)^5$ flavor currents of the SM
\begin{equation}\label{eq:lag_mass_basis}
    \mathcal{L} \supset \frac{\partial_\mu a}{f_a} \sum_\psi c^a_{\psi L} (j^a_L)^\mu + c^a_{\psi R} (j^a_R)^\mu\;,
\end{equation}
with $c^a_{uL} = c^a_{dL}$ and $c^a_{e L} = c^a_{\nu L}$ when Yukawa interactions vanish. The above currents individually satisfy Ward identities and are gauge invariant under the SM gauge groups.

In this work, we consider only the interactions of $SU(3)_c$ for simplicity, and the leptonic part will be ignored in further discussions. Crucially, the currents are $SU(3)_c$ gauge invariant, and their Ward identities will hold under renormalization by strong interactions to all orders. The anomalous breaking of chiral global symmetries at the quantum level is taken into account in the Ward identities through the anomaly term $G\tilde{G}$, which does not vanish as $\text{Tr}(t^a)\neq 0$. Hence, the ALP operator $\mathcal{O}_G$ can also be renormalized using Ward identities. A final point is that the dimension-five terms in the full ALP Lagrangian Eq.~\eqref{eq:UVlag} are \textit{not} invariant under global symmetry because the $U(3)$ generators $t^a$ do not commute with arbitrary Wilson coefficient matrices. However, this does not affect the above renormalization arguments, since only mass-dimension four SM gauge interactions are involved in renormalizing the flavor currents. 

Our strategy can now be summarized transparently. Renormalization of mass-dimension five ALP EFT operators is achieved by renormalizing the SM currents they contain. In the massless fermion limit, these are the Noether currents of the $U(3)$ chiral global symmetries of the renormalizable Lagrangian and satisfy certain Ward identities. The BMHV scheme enables a robust derivation of dimensionally-regularized identities and subsequent renormalization. In the next section, we present these arguments in a more formal manner.

\section{Flavor Current Ward Identities in BMHV}\label{sec:WIs}
In this section, we will deduce bare and renormalized Ward identities (WI) in the BMHV scheme for the chiral flavor currents discussed earlier. The theoretical foundations are essentially based on~\cite{Collins:1984xc}, but we will motivate the definitions of $d$-dimensional quantities in greater detail. 

\subsection{Four-dimensional Lagrangian}
We start with a renormalizable Lagrangian consisting of massless chiral fermion multiplets $\{\psi_L, \psi_R \}$ that are gauge invariant under $SU(3)_c$. For this discussion, the gauge boson kinetic and gauge-fixing terms are irrelevant and will not appear subsequently. Working in $d=4$ dimensions, the Lagrangian with bare fields reads 
\begin{equation}\label{eq:lagrangian}
    \mathcal{L}_{d=4} = \bar{\psi}_L i\slashed{D} \psi_L + \bar{\psi}_R i\slashed{D} \psi_R\;.
\end{equation}
We reiterate that chiral fields form a triplet in the generation space, e.g. $\psi_{\text{up}} = \{u,c,t\}$. The corresponding indices will be kept implicit. The covariant derivative for $SU(3)_c$ is diagonal in the generation space $ (\slashed{D})_{ij} \equiv \delta_{ij} (\slashed{\partial} - ig_s\slashed{G})$, where $G_\mu \equiv G^a_\mu T^a$ is the corresponding $SU(3)_c$ gauge boson field. Expressed with projection operators, the Lagrangian becomes 
\begin{equation}\label{eq:lag4}
    \mathcal{L}_{d=4} =  \bar{\psi} P_R i\slashed{D} P_L \psi + \bar{\psi} P_L i\slashed{D} P_R \psi \;.
\end{equation}
At this stage, one can think of manipulating the Dirac matrices and projectors by the usual $d=4$ algebra and write $\mathcal{L}_{d=4} = \bar{\psi}i\slashed{D}\psi$. However, we avoid this for reasons that will become clear in Sect.~\ref{subsec:d-dim}. The $U(3)$ global chiral transformations acting in the generation space defined in Eq.~\eqref{eq:chiraltransformation} give the following Noether currents
\begin{equation}\label{eq:noetherdefinition}
    j^{a\mu}_{L/R} \equiv - \frac{\partial \mathcal{L}_{d=4}}{\partial (\partial_\mu \psi)} \frac{\delta \psi}{\delta \alpha^a_{L/R}}  \;.
\end{equation}
The derivative with respect to $\partial_\mu \bar{\psi}$ will give zero and is omitted. Inserting the Lagrangian form of Eq.~\eqref{eq:lag4} in Eq.~\eqref{eq:noetherdefinition}, we find the currents given in Eq.~\eqref{eq:currentsLR}.

\subsection{Continuation to \texorpdfstring{$d$-dim}{d-dim}}\label{subsec:d-dim}
In the framework of dimensional regularization, we must define $d$-dimensional currents and Lagrangian for renormalization. We define the former directly from Eq.~\eqref{eq:currentsLR} by treating the $P_{L/R}$ and $\gamma^\mu$ that appear as $d$-dimensional. The specific positions of $P_{L/R}$ in Eq.~\eqref{eq:currentsLR} are a direct consequence of starting with chiral fermions, as in Eq.~\eqref{eq:lagrangian}. This fits naturally with the requirement of having Hermitian chiral currents in schemes with non-anticommuting $\gamma_5$. Had we chosen $\mathcal{L}_{d=4} = \bar{\psi}i\slashed{D}\psi$, keeping the rest of the procedure the same, we would have obtained non-Hermitian currents $j^{a\mu}_{L/R} = \bar{\psi} \gamma^\mu P_{L/R}t^a\psi$ in $d\neq 4$. Thus, starting with chiral fermions, even in a vector gauge theory, is a useful prescription to obtain Hermitian currents in $d$-dimensions directly. 

For the Lagrangian, the kinetic part must be made $d$-dimensional to obtain $d$-dimensional propagators of Dirac fields for dimensionally regularized integrals. There exist multiple ways to achieve this that are equivalent in four dimensions, but inequivalent in arbitrary $d$ dimensions due to non-anticommuting $\gamma_5$~\cite{Belusca-Maito:2023wah}. Our guiding principle will again be to maintain hermiticity in the $d$-dimensions using the BMHV algebra (Appendix~\ref{app:HValgebra}).

\begin{enumerate}
    \item We begin by separating the kinetic and interaction terms in the four-dimensional Lagrangian Eq.~\eqref{eq:lag4} and continue the dimensions to $d\neq 4$
\begin{eqnarray}\label{eq:symmetrizedL}
    \mathcal{L}_{d=4} &=& \mathcal{L}_{\text{kin}} + \mathcal{L}_{\text{int}} \to \mathcal{L}_{d\neq 4}\;, \nonumber \\
   \mathcal{L}_{d\neq 4} &=& i \bar{\psi} (P_R\slashed{\partial}P_L + P_L\slashed{\partial}P_R) \psi + \bar{\psi}( P_R\slashed{G}P_L + P_L\slashed{G}P_R)\psi\;,\nonumber\\
   &=& i\bar{\psi} \widetilde{\slashed{\partial}} \psi + \bar{\psi} \widetilde{\slashed{G}} \psi\;.
\end{eqnarray}
In the last step, we have used the BMHV algebra to simplify the expressions. However, these unambiguous steps do not produce the correct regularized Lagrangian because the kinetic part $\bar{\psi} i \tilde{\slashed{\partial}}\psi $ is in the purely four-dimensional subspace. This leads to unregularized propagators in momentum space, defeating the purpose of dimensionally regularizing divergent loop integrals~\cite{Jegerlehner:2000dz, Belusca-Maito:2023wah}.

\item Next, to cure this problem, we algebraically manipulate $\mathcal{L}_{d\neq 4}$ with BMHV algebra to split it into a formal $d$-dim part and an evanescent  $\epsilon$-dim part.
\begin{equation}
   \mathcal{L}_{d\neq 4} =  \mathcal{L}_{d-\text{dim}} + \mathcal{L}_{\text{ev}}\;.
\end{equation}
The splitting should be performed such that the kinetic part in $\mathcal{L}_{d-\text{dim}}$ gives regularized $d$-dim propagators and $\mathcal{L}_{\text{ev}}$ contains only objects in the $\epsilon$-dim subspace that vanish as $d\to 4$. For the kinetic part,  
\begin{eqnarray}
    \mathcal{L}_{\text{kin},\ d\neq 4} &=& \mathcal{L}_{\text{kin, d-dim}} + \mathcal{L}_{\text{kin-ev}}\;, \nonumber \\
    &=& \bar{\psi} i \slashed{\partial} \psi + \mathcal{L}_{\text{kin-ev}} \;, \quad   \mathcal{L}_{\text{kin-ev}} = -\bar{\psi} i \widehat{\slashed{\partial}} \psi\;.
\end{eqnarray}
Note that the evanescent part $\mathcal{L}_{\text{ev}}$ is formally on a different footing from the $d$-dim Lagrangian and not used, for example, in perturbation theory Feynman rules; the effect of this splitting nevertheless enters into the breaking of usual Ward identities by evanescent operators, as will be discussed later.
\item  Similarly for the gauge interaction part,
\begin{equation}\label{eq:ddim_interaction}
\begin{aligned}
    \mathcal{L}_{\text{int},\ d\neq 4}  &= \bar{\psi}( P_R\slashed{G}P_L + P_L\slashed{G}P_R)\psi= \mathcal{L}_{\text{int, d-dim}} + \mathcal{L}_{\text{int, ev}}\;, \\
    &=   \bar{\psi} \slashed{G}\psi + \mathcal{L}_{\text{int-ev}}\;,\quad  \mathcal{L}_{\text{int-ev}} = -  \bar\psi \slashed{\widehat{G}} \psi\;.
\end{aligned}
\end{equation}
\end{enumerate}

The above procedure is fairly straightforward because we are dealing with a vector gauge theory that does not distinguish between the left-handed and right-handed fermions. This fails for a chiral gauge theory, as their wavefunction renormalizations differ. In that case, the splitting mentioned in steps 2 and 3 can be done in different and inequivalent ways~\cite{Belusca-Maito:2023wah}. Individual maintenance of hermiticity for the left and right chiral kinetic terms restricts the form of $\mathcal{L}_{\text{kin, d-dim}}$. For the interaction term in Eq.~\eqref{eq:ddim_interaction}, the alternate choice of taking a symmetrized vertex $\widetilde{\slashed{G}} = \widetilde{\gamma}^\mu G_\mu $ as $\mathcal{L}_{\text{int, d-dim}}$ is not forbidden by the requirement of regularized loop integrals. However, this apparently simpler choice does not yield a manifestly gauge-invariant $d$-dimensional Lagrangian. As a result, the fermion self-energy calculations themselves acquire non-trivial evanescent aspects. This is in contrast to the usual prescription of taking a symmetrized left-handed $W$-boson vertex for the chiral gauge theory of weak interactions within the BMHV scheme~\cite{Ferrari:1994ct, Buras:1989xd, Buras:2020xsm}. Such distinct aspects of chiral gauge theories in the BMHV scheme require a dedicated future study.

\subsection{Ward Identities}
In quantum field theory, it is important to explore the consequences of the classical properties of the action on Green's functions. These are crucial for the mathematical consistency of the renormalized theory and provide a link to physical observables. Focusing on internal symmetries of the classical action with a conserved current $j^\mu(x)$, there exist relations between Green functions known as Ward identities. They encode classical current conservation as well as the quantum equal-time commutation relation $[j^0(x), \psi(y)] = -i\Delta \psi(x)\delta^{(3)}(x-y)$. We use fermions for concreteness, and $\Delta \psi$ denotes the change in the fermion field under the symmetry transformation, with the transformation parameter extracted. We write $\partial_\mu j^\mu = \mathcal{O}_{\text{eom}}$, where $\mathcal{O}_{\text{eom}}$ vanishes on-shell. It can be obtained by calculating $\partial_\mu j^\mu$ algebraically without using the equations of motion in $d=4$. This can be inserted into Green functions, yielding Ward identities. For example,
\begin{equation}\label{eq:Ward}
\begin{aligned}
        \frac{\partial}{\partial x^\mu}\langle 0 |T j^\mu(x) \psi(x_1) \bar{\psi}(x_2)|0\rangle  =  \langle 0 |T\; \mathcal{O}_{\text{eom}}(x) \psi(x_1) \bar{\psi}(x_2)|0\rangle\;. 
\end{aligned}
\end{equation}
The equation of motion operator is defined to give contact terms\footnote{ Therefore, $\mathcal{O}_{\text{eom}}$ is manifestly finite and renormalizes itself~\cite{Collins:1984xc, Bos:1992nd}. However, this has complications when working with chiral gauge interactions, which will be discussed in the future.} which arise as the time derivative on the left side is taken inside time-ordering~\cite{Collins:1984xc}
\begin{eqnarray}
   \langle 0 |T\; \mathcal{O}_{\text{eom}}(x) \psi(x_1) \bar{\psi}(x_2)|0\rangle = &-& i \delta^{(4)}(x-x_1) \langle 0 | T \Delta \psi(x) \bar{\psi}(x_2) |0\rangle  \nonumber \\
   &-& i \delta^{(4)}(x-x_2) \langle 0 | T \Delta \bar{\psi}(x) \psi(x_1) |0\rangle\;.
\end{eqnarray}
The equation of motion operator is defined independently via the path integral method
\begin{equation}\label{eq:Oeom}
    \mathcal{O}_{\text{eom}} \equiv \underbrace{\frac{\delta \mathcal{L}}{\delta \psi}}_{\equiv \bar{D}_{\text{eom}}} \Delta \psi + \Delta \bar{\psi}  \underbrace{\frac{\delta \mathcal{L}}{\delta \bar{\psi}}}_{\equiv D_{\text{eom}}}\;.
\end{equation}
When no explicit symmetry-breaking effects are present, this definition and the evaluation of $\partial_\mu j^\mu$ off-shell are the same. However, symmetry-breaking effects add additional terms on the right side of Eq.~\eqref{eq:Ward}. A specific type of symmetry-breaking occurs when one regularizes Ward identities using dimensional regularization with a non-anticommuting $\gamma_5$ such as in the BMHV scheme. This is the topic of the subsequent sections.

\subsubsection{Bare Ward Identities and Evanescent Operators}
In the BMHV scheme, the four-dimensional Ward identities that encode current conservation are generally violated away from four dimensions. This happens due to the presence of evanescent operators linked to the breaking of the chiral symmetry in $d\neq 4$ dimensions~\cite{Belusca-Maito:2023wah}. We explicitly deduce this for the Ward identities of our chiral flavor currents in this section.
 
To derive $d$-dimensional Ward identities, we compute the divergence of the $d$-dimensional currents and identify the $d$-dimensional equation of motion operator from Eq.~\eqref{eq:Oeom}. The remaining terms breaking the naive Ward identities will turn out to be the evanescent operators. For our chiral global symmetry, the equation of motion operator is
\begin{equation}\label{eq:Oeom_ddim}
    (\mathcal{O}_{\text{eom}}^a)_{L/R} \equiv \underbrace{\frac{\delta \mathcal{L}}{\delta \psi}}_{\equiv \bar{D}_{\text{eom}}} \left(it^a P_{L/R} \psi\right) -\left(i \bar{\psi}  P_{R/L}\ t^a\right) \underbrace{\frac{\delta \mathcal{L}}{\delta \bar{\psi}}}_{\equiv D_{\text{eom}}}\;. 
\end{equation}
Using $\mathcal{L}_{d-\text{dim}}$ in the above equation, we find the following.
\begin{equation}
\begin{aligned}
& (\mathcal{O}_{\text{eom}}^a)_{L/R}  = \bar{D}_{\text{eom}} \left(it^a P_{L/R} \psi\right) - \left(i\bar{\psi} P_{R/L}t^a\right) D_{\text{eom}} \;, \\
       & \bar{D}_{\text{eom}} = \bar{\psi} \slashed{G} -  \partial_\mu \bar{\psi} i \gamma^\mu \;, \quad D_{\text{eom}} = i\gamma^\mu \partial_\mu \psi  +  \slashed{G} \psi\;.
\end{aligned}
\end{equation}
Hence, the $d$-dimensional equation of motion operators are hermitian. Evaluating $\partial . j_{L/R}^a$  and expressing the results in terms of $(\mathcal{O}^a_{\text{eom}})_{L/R}$ gives
\begin{equation}\label{eq:WILR}
\begin{aligned}
            \partial . j^a_{L/R} &= (\mathcal{O}^a_{\text{eom}})_{L/R} + ( - \partial_\mu \bar{\psi}\widehat{\gamma}^\mu P_{L/R} t^a \psi -  \bar{\psi} \widehat{\gamma}^\mu P_{R/L} t^a \partial_\mu \psi  \pm i \bar{\psi} \slashed{\widehat{G}}\gamma_5 t^a \psi )\;,  \\
            &\equiv  (\mathcal{O}^a_{\text{eom}})_{L/R} + (\mathcal{O}^a_E)_{L/R}\;.
\end{aligned}
\end{equation}
The evanescent operators $(\mathcal{O}^a_E)_{L/R}$ appearing in the bare $d$-dimensional Ward identities seemingly violate their four-dimensional form. Containing entities belonging to the $\epsilon$-dimensional subspace, they vanish at tree-level as $d\to 4$. We note that the chiral transformations of Eq.~\eqref{eq:chiraltransformation} can also be written as $(\alpha_L^a P_L + \alpha_R^a P_R)t^a  = (\alpha_V^a + \alpha_A^a \gamma_5)t^a$ where $\alpha_{V/A}^a \equiv (\alpha^a_R \pm \alpha^a_L)/2$, giving vector and axial-vector currents
\begin{equation}\label{eq:vector_axial_currents}
    j^{a\mu}_V =   j^{a\mu}_R + j^{a\mu}_L =  \bar{\psi}(\gamma^\mu - \widehat{\gamma}^\mu)t^a \psi \;, \quad j^{a\mu}_A=  j^{a\mu}_R - j^{a\mu}_L = \frac{1}{2}\bar{\psi}( \gamma^\mu \gamma_5 - \gamma_5 \gamma^\mu)t^a \psi \;.
\end{equation}
and the corresponding equation of motion operators;
\begin{equation}
    (\mathcal{O}^a_{\text{eom}})_V = i\bar{D}_{\text{eom}} t^a \psi - i
    \bar{\psi} t^a D_{\text{eom}} \;, \quad  (\mathcal{O}^a_{\text{eom}})_A = i\bar{D}_{\text{eom}} t^a  \gamma_5 \psi + i
    \bar{\psi} \gamma_5 t^a D_{\text{eom}}\;.
\end{equation}
This change of basis offers practical advantages for purely vector gauge interactions,  since they conserve parity, and the vector and axial-vector currents do not mix under renormalization in \textit{any} dimensional regularization scheme (see the end of Sect.~\ref{sec:renWI}). In terms of vector and axial vector currents, the bare $d$-dimensional identities are
\begin{equation}\label{eq:bareWardoperatorlevel}
\begin{aligned}
 \partial . j^a_V &= i\bar{D}_{\text{eom}} t^a \psi + \bar{\psi}\ t^a (-iD_{\text{eom}}) - \partial_\mu \bar{\psi} \widehat{\gamma}^\mu t^a \psi -  \bar{\psi}\widehat{\gamma}^\mu t^a \partial_\mu \psi\;, \\
     \partial . j^a_A &= i\bar{D}_{\text{eom}} t^a \gamma_5 \psi + \bar{\psi}\gamma_5 t^a (iD_{\text{eom}}) -  \partial_\mu \bar{\psi} \widehat{\gamma}^\mu t^a \gamma_5 \psi + \bar{\psi}\widehat{\gamma}^\mu  t^a \gamma_5 \partial_\mu \psi  -2i\bar{\psi} \slashed{\widehat{G}} t^a \gamma_5 \psi\;.
\end{aligned}
\end{equation}
The evanescent structure in the vector current equation is a total derivative and does not contribute to perturbative matrix elements (see Sect.~\ref{sec:fin_ren}). Contrastingly, the evanescent structure in the axial-vector current leads to non-trivial results discussed in the following sections. We arrive at the following bare Ward identities with external fermions
\begin{equation}\label{eq:bareWard}
\begin{aligned}
 \frac{\partial}{\partial z^\mu}\langle 0|T\{j^{a\mu}_V(z)\  \psi \bar{\psi} \}|0\rangle &= \langle 0|T \{(\mathcal{O}^a_{\text{eom}})_V(z)\psi  \bar{\psi} \} |0\rangle\;,\\
    \frac{\partial}{\partial z^\mu}\langle 0|T\{j^{a\mu}_A(z)\  \psi \bar{\psi} \}|0\rangle &= \langle 0|T \{(\mathcal{O}^a_{\text{eom}})_A(z)  \psi  \bar{\psi}\} |0\rangle + \langle 0|T\{\mathcal{O}^a_E(z) \psi  \bar{\psi}\} |0\rangle \;.
\end{aligned}
\end{equation}
where $\mathcal{O}^a_E \equiv  -  \partial_\mu \bar{\psi} \widehat{\gamma}^\mu t^a \gamma_5 \psi + \bar{\psi}\widehat{\gamma}^\mu  t^a \gamma_5 \partial_\mu \psi -2i\bar{\psi} \slashed{\widehat{G}} t^a \gamma_5 \psi$ is the evanescent operator. In Sec.~\ref{sec:axialresults}, these identities will be shown to hold up to $\mathcal{O}(\alpha_s^2)$.

\subsubsection{Renormalized Ward Identities}\label{sec:renWI}
The Ward identities in Eq.~\eqref{eq:bareWard} are in $d$-dimensions and contain bare operators and fields. The next step is to obtain Ward identities with renormalized quantities. We denote fermion field renormalization and composite operator renormalization as~\footnote{Here we chose an opposite convention for operator renormalization with respect to our previous work~\cite{Bisht:2024hbs} as it led to simpler calculations in Sec.~\ref{sec:ADM}}
\begin{equation}\label{eq:notationrenormalization}
    \psi = \mathcal{R}^{1/2}_{\psi} \psi^{\text{ren}}\;,\quad  \mathcal{O} = \hat{\mathcal{Z}}^{-1} [\mathcal{O}]\;,
\end{equation}
where the quantities on the left are bare, and the right ones are renormalized. The hat $\hat{}$ denotes matrix structure, as we work with multiple operators that can mix under renormalization. We employ dimensional regularization in $d=4-2\epsilon$ and $\overline{\text{MS}}$, so the renormalization constants $\hat{\mathcal{R}}, \hat{\mathcal{Z}}$ contain poles in $1/\epsilon$. They also have a perturbative expansion in the gauge coupling(s).  
It follows that bare and renormalized amputated Green's functions with operator insertion are related as
\begin{equation}\label{eq:green_fn_renorm}
    \langle \mathcal{O} \rangle = \hat{\mathcal{R}}^{-1} \hat{\mathcal{Z}}^{-1}\langle [\mathcal{O}]\rangle = \hat{\chi}\langle [\mathcal{O}]\rangle \;, \quad \hat{\chi} \equiv \hat{\mathcal{R}}^{-1} \hat{\mathcal{Z}}^{-1} \;.
\end{equation}
The complete renormalization constant $\hat{\chi}$ absorbs all poles, making $\langle [\mathcal{O}]\rangle$ finite. Then elements of $\hat{\mathcal{Z}}$ are found from the poles in $\hat{\mathcal{R}}$ and $\hat{\chi}$. In general, $\hat{\mathcal{R}}$ is a non-diagonal matrix, but for purely $\mathcal{O}(\alpha_s)$ interactions, it is diagonal because different operators never mix by wavefunction renormalization. Also, $\mathcal{R}_\psi$ in Eq.~\eqref{eq:notationrenormalization} is independent of chirality, generation, or quark species (up or down).

For the vector Ward identity, the right side only contains $(\mathcal{O}^a_{\text{eom}})_V$. This operator is manifestly finite~\cite{Bos:1992nd, Collins:1984xc}, i.e. the divergences in its matrix elements are canceled by the associated wavefunction renormalization to all orders in $\alpha_s$. It also does not mix under renormalization with the vector current. Thus,  $(\mathcal{O}^a_{\text{eom}})_V = [(\mathcal{O}^a_{\text{eom}})_V]$, with $\mathcal{Z}_{\text{eom}} = 1$. Taking $[j^{a\mu}_V] = \mathcal{Z}_V j^{a\mu}_V$, we find that the vector current is not renormalized at any $\alpha_s$ order. 
\begin{equation}
 \partial . j^a_V = (\mathcal{O}_{\text{eom}}^a)_V \Rightarrow   \mathcal{Z}_V^{-1} \partial. [j^a_V] = [(\mathcal{O}^a_{\text{eom}})_V] \Rightarrow \mathcal{Z}_V = 1\;.
\end{equation}
More interestingly, the axial Ward identity containing the evanescent operator $\mathcal{O}^a_E$ leads to non-trivial operator renormalization. Firstly, the axial current can only mix with itself, but the evanescent operator can mix with the current. Defining
\begin{equation}
    [j^{a\mu}_A] = \mathcal{Z}_A j^{a\mu}_A \;, \quad  [\mathcal{O}_E^a] = \mathcal{Z}_E \mathcal{O}_E^a  + \mathcal{Z}_{EA} \partial . j^a_A \;,
\end{equation}
using the finiteness of the axial equation of motion operator, we derive
\begin{equation}
\begin{aligned}
\partial . j^a_A &= (\mathcal{O}^a_{\text{eom}})_A + \mathcal{O}^a_E\;, \\
\Rightarrow [(\mathcal{O}^a_{\text{eom}})_A] &=   -\mathcal{Z}_{E}^{-1}[\mathcal{O}^a_E]  +(Z_E^{-1}Z_{EA} + 1) Z_A^{-1} [\partial. j^a_A]\;. 
\end{aligned}
\end{equation}
The renormalized operators on the right are linearly independent and finite, so their coefficients must be individually finite to yield the finite left side~\cite{Collins:1984xc}. Thus,  
\begin{equation}
  \mathcal{Z}_E = 1\;, \quad  \mathcal{Z}_A = 1 + \mathcal{Z}_{EA} \;,
\end{equation}
to all orders in $\alpha_s$. The renormalized vector and axial-vector Ward identities are
\begin{equation}\label{eq:renorm_WI}
\begin{aligned}
 [\partial .j^a_V] = [(\mathcal{O}^a_{\text{eom}})_V] \;,\quad [\partial .j^a_A] = [(\mathcal{O}^a_{\text{eom}})_A] + [\mathcal{O}^a_E]\;.
\end{aligned}
\end{equation}
These Ward identities and their consequences for the anomalous dimensions of the operators will be shown to hold explicitly up to two-loops $\mathcal{O}(\alpha_s^2)$ in the remaining sections. Before moving on to these primary results, we make a few important remarks.
\begin{enumerate}
    \item The Ward identities of \textit{both} left and right chiral currents contain evanescent operators (see Eq.~\eqref{eq:WILR}), and since the purely vector gluon vertex contains both chiralities $\mathds{1} = P_L + P_R$, there is mixing of opposite chirality operators under renormalization. These complications have been avoided by working in the $V-A$ basis. 

    \item In NDR, the regularized currents and Lagrangian carry the same form as in $d=4$. As no evanescent subspace is defined, evanescent operators do not arise in the regularized Ward identities. The presence of evanescent operators is manifestly related to nonanticommuting $\gamma_5$. However, in studies involving four-fermi operators, the relevant Fierz identities are broken simply by continuation to $d\neq 4$, irrespective of the particular dim-reg scheme. Both NDR and BMHV then contain Fierz-breaking evanescent operators~\cite{Buras:1989xd}.
    
    \item For the vector WI, no $\gamma_5$ is involved and both BMHV and NDR give identical results, which we verified explicitly up to two-loops. For the axial WI, qualitative and quantitative differences occur in the two schemes. In BMHV, the evanescent operator is the source of the $G\widetilde{G}$ axial anomaly. We explicitly show this in Sec.~\ref{sec:ADM}. Whereas, in the NDR scheme, there is no direct way to infer the axial anomaly~\cite{Bhattacharya:2015rsa}. Still, a one-loop calculation, forbidding trace cyclicity, enables a derivation~\cite{Korner:1991sx}. However, in Appendix~\ref{app:NDR}, we point out that the canonical axial WI with the current and $G\widetilde{G}$ does not hold at the \textit{two-loop} level in NDR without some ad hoc finite renormalization.
\end{enumerate}
Our treatment based on~\cite{Collins:1984xc} offers a mathematically equivalent but conceptually cleaner treatment of evanescent operators. In the literature, one starts with some $d$-dimensional Lagrangian and finds suitable evanescent counterterms by imposing symmetry requirements~\cite{Belusca-Maito:2023wah}. In the present approach, the necessary evanescent operators emerge algebraically, and the regularized Ward identity, including the evanescent operator, is always satisfied.  

\section{Ward Identities for Non-singlet Axial Current}\label{sec:axialresults}
In this section, we explicitly demonstrate that the Ward identity remains satisfied up to two-loop order in the presence of evanescent contributions, even in dimensions $d \neq 4$. Owing to the additional global $U(3)^5$ flavor symmetry of the massless Standard Model in generation space, under which quarks transform non-trivially only under color, we restrict our analysis to purely gluonic corrections.
\begin{figure}[H]
    \centering
\begin{tikzpicture}
\begin{feynman}
    \vertex (a);
    \vertex [right= of a, blob, minimum size=0.6cm, inner sep=0pt] (b) {}; 
    \vertex [right = 1.8cm of b] (c);

    \diagram* {
      (a) -- [fermion, edge label=$p$] (b) -- [fermion, edge label=$p'$] (c),
    };

    \node [above=0.8cm of b] {$\;\Big\downarrow  \ell$};
\end{feynman}
\end{tikzpicture}
\caption{Blob represents the tree-level operator insertion.}
\label{fig:operator_insertion}
\end{figure}
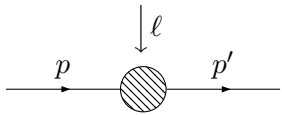

Let us take one of the $U(3)$ symmetries in the quark generation space, the corresponding non-singlet axial current in Eq.~\eqref{eq:vector_axial_currents},  
\begin{equation}
    j^a_{A\mu} (z) = \frac{1}{2}  \bar\psi_i(z) \qty[\gamma^\mu, \gamma_5] t^a_{ij} \psi_j(z)\;.
\end{equation}
The vertex function $\Gamma_A^\mu$ for this current is defined as
\begin{equation}\label{eq:vertex_fn}
    S_F(p')_{im}\Gamma^a_{A\mu}(p',p)_{mn} S_F(p)_{nj} = \int d^4x\, d^4y\,d^4z \,e^{i(p'\cdot x - p\cdot y - l\cdot z)} \expval{T\qty{j^a_{A\mu}(z)\psi_i(x)\bar\psi_j(y)}}\;.
\end{equation}
Similarly, we can define the vertex function for the evanescent operator by simply replacing $j^{a\mu}_A$ with $\mathcal O^a_E$. The momentum directions of the external fields are represented schematically in Fig.~\ref{fig:operator_insertion}. From Eq.~\eqref{eq:bareWard}, we obtain the bare Ward-Takahashi identity in momentum space,
\begin{equation}\label{eq:ward_momentum}
    -i(p'-p)^\mu \Gamma^a_{A\mu}(p,p')_{ij} = t^a_{ik} \gamma_5 S_F^{-1}(p)_{kj} + S_F^{-1}(p')_{ik}  \gamma_5 t^a_{kj} - \Gamma^a_{E}(p,p')_{ij}\;.
\end{equation}
In the following expressions, we will suppress the generation indices. It is convenient to expand the bare propagator and vertex functions in $\alpha_s$,
\begin{align}
     S_F^{-1}(p) & = -i\slashed p  \qty[1 + \qty(-\delta \mathcal R_\psi^{(1)}  +  \Sigma^{(1)} )\frac{\alpha_s}{4\pi} + \qty(\delta \mathcal R_\psi^{(2)} + \Sigma^{(2)}) \left(\frac{\alpha_s}{4\pi}\right)^2]\;,  \\
    \Gamma^\mu_A(p,p') & = \widetilde\gamma^\mu  \gamma_5 \qty[1 +  \qty(\delta \chi_A^{(1)} + F^{(1)}) \frac{\alpha_s}{4\pi} +  \qty(\delta \chi_A^{(2)} + F^{(2)}) \left(\frac{\alpha_s}{4\pi}\right)^2]\;,  \\
     \Gamma_{E}(p,p') & = -i (\slashed 
     {\widehat p} + \slashed{\widehat p'}) \gamma_5  \qty[1 +   \qty(\delta \chi_{E}^{(1)} + E^{(1)}) \frac{\alpha_s}{4\pi} +  \qty(\delta  \chi_{E}^{(2)} + E^{(2)}) \left(\frac{\alpha_s}{4\pi}\right)^2] \nonumber\\
     & \quad - i (\slashed {\widetilde p'} - \slashed {\widetilde p})  \gamma_5  \qty[\qty(\delta \chi_{EA}^{(1)}  + K^{(1)}) \frac{\alpha_s}{4\pi} +  \qty(\delta \chi_{EA}^{(2)} + K^{(2)})  \left(\frac{\alpha_s}{4\pi}\right)^2] \;.
\end{align} 
Using Eq.~\eqref{eq:ward_momentum}, at the tree level, the Ward identity is trivially satisfied even in $d\neq 4$. At the one-loop level, the poles are
\begin{equation}\label{eq:poles_1l}
\begin{aligned}
    \delta \mathcal R_\psi^{(1)}  =  - \frac{C_F}{\epsilon} I,  \quad \delta \chi_{EA}^{(1)} = 0\;, \\
    \quad \delta \chi_A^{(1)} = \delta \chi_E^{(1)} = - \delta \mathcal R_\psi^{(1)} \odot t^a,
    \end{aligned}
\end{equation}
 where $\odot$ denotes the Hadamard product, $I$ being the identity matrix $3\times3$, and now the finite pieces are
\begin{equation}\label{eq:finite_pieces_1l}
\begin{aligned}
        \Sigma^{(1)} & =  C_F\qty[1 + \log(\frac{\mu^2}{-p^2})] I,\\ F^{(1)} & =  (\Sigma^{(1)}  + 4 C_F I) \odot t^a\;, \\
        E^{(1)} & = \Sigma^{(1)} \odot t^a,\\ 
        K^{(1)} & = - 4 C_F I\odot t^a\;.
\end{aligned}
\end{equation}
These satisfy the renormalization conditions and the renormalized Ward identities, respectively. Examining the operator structure, we observe that, to $\alpha_s^2$ orders in perturbation theory, the physical current does not mix with other operators under renormalization. In contrast, the evanescent operator exhibits non-trivial mixing into the four-dimensional current beyond leading order. However, this mixing does not correspond to the correct finite projection onto the physical current, an issue that will be addressed in the following section. We further corroborate our analysis through an explicit two-loop calculation, demonstrating that both the pole and finite contributions are consistently reproduced.
\begin{equation}\label{eq:poles_2l}
\begin{aligned}
    \delta \mathcal R_\psi^{(2)} & = - \qty[\frac{1}{\epsilon^2}C_F(2C_A + C_F) + \frac{1}{4\epsilon}C_F(-17C_A+3C_F + 4n_F T_F)] I\;,  \\
    \delta \chi_A^{(2)} & = \qty[ - \delta \mathcal R _\psi^{(2)} - \frac{2}{3\epsilon} C_F(11 C_A - 4 n_F T_F) I] \odot t^a - \frac{6}{\epsilon}C_F T_F\Tr(t^a) I\;,\\
    \delta \chi_E^{(2)} & = - \delta \mathcal R_\psi^{(2)} \odot t^a, \quad\delta \chi_{EA}^{(2)} = \delta \chi_{A}^{(2)}\;,\\
\end{aligned}
\end{equation}
\begin{equation}\label{eq:finite_pieces_2l}
\begin{aligned}
    \Sigma^{(2)} & =  \frac{C_F}{24}\bigg[-15C_F - 314 C_A - 84 n_F T_F  + 12(21 C_A - C_F - 4 n_F T_F) \log(\frac{\mu^2}{-p^2}) \\
     & \qquad + 12(2C_A+ C_F) \log^2\qty(\frac{\mu^2}{-p^2}) + f\qty(\psi)\bigg] I\;, \\
    F^{(2)} & =  \bigg[\Sigma^{(2)}  +  \frac{C_F}{9}(107 C_A - 18 C_F - 4 n_F T_F) I  \\ 
    & \qquad + 4C_F (C_F - 3 T_F)\log(\frac{\mu^2}{-p^2})I\bigg] \odot t^a - 27 C_F T_F\Tr(t^a) I\;, \\
    E^{(2)} & = \Sigma^{(2)} \odot t^a, \\
    K^{(2)} & = \qty[\frac{C_F}{9}(107 C_A - 18 C_F - 4 n_F T_F) + 4C_F(C_F - 3 T_F)\log(\frac{\mu^2}{-p^2}) ] I \odot t^a \\
    & \qquad - 27 C_F T_F\Tr(t^a) I\;,
\end{aligned}
\end{equation}
where 
\begin{equation*}
    f(\psi) =  4C_A\qty( 7\psi^{(2)}(1) - 16\psi^{(2)}(2) + 27 \psi^{(2)}(3))\,.
\end{equation*}
The $\psi^{(n)}(z)$ refers to the polygamma function, defined as $\dv*[n+1]{\log\Gamma(z)}{z}$, where $\Gamma(z)$ is the usual gamma function. Since the generators corresponding to the $U(3)$ flavor symmetry are not traceless, the triangle diagrams generated by inserting the non-singlet current at higher loops are non-zero and give the $\Tr(t^a)$ part, which means the non-singlet current mixes with a singlet current.
In Appendix~\ref{app:diagrams}, we outline the computational procedure used for the generation and evaluation of the two-loop diagrams.

\section{Finite Renormalization and Anomalous Dimension Matrix}\label{sec:ADM}
 In the above section, we explicitly verified the regularized Ward identity that contains the evanescent operator up to two loops. However, knowing that the evanescent operator is a scheme-dependent object, one is motivated to derive a four-dimensional Ward identity containing only physical operators (i.e., whose vertex structures are non-evanescent). This should match the form of the canonical and scheme-independent anomalous axial Ward identity~\cite{Adler:1969er, Kodaira:1979pa, Larin:1993tq}. In this section, we show that this can be achieved as the evanescent operator can be projected onto the basis of physical operators through its loop-level matrix elements. This will allow us to express the Ward identity purely in terms of non-evanescent operators. Moreover, this leads to the identification of the physical axial current, related to the $\overline{\text{MS}}$ current by finite renormalization. This procedure to determine the finite renormalization in the BMHV scheme is a direct first-principles approach based on~\cite{Collins:1984xc} but implemented at the two-loop level for the first time in this work. It offers an alternative to the standard method developed in~\cite{Larin:1993tq}. Additionally, we also derive the anomalous dimension matrix elements for the physical currents and the anomaly operator $G\widetilde{G}$ in this section, and verify that they satisfy the relations dictated by the Ward identity.

\subsection{Finite Renormalization}\label{sec:fin_ren}
A renormalized evanescent operator can be expressed as a linear combination of non-evanescent operators, constrained by mass dimension and symmetry considerations \cite{Collins:1984xc}. In the present case, the relevant non-evanescent operator basis consists of $\partial \cdot j, G\widetilde{G}$, and the equation-of-motion operator $\mathcal{O}_{\text{eom}}$. However, employing the renormalized axial Ward identity in Eq.~\eqref{eq:renorm_WI}, the operator $\mathcal{O}_{\text{eom}}$ can be eliminated in favor of $\mathcal{O}_E$ and $\partial \cdot j$, which amounts to a redefinition of the corresponding coefficients. Furthermore, since $\mathcal{O}_E$ appearing in the Ward identity carries a $U(3)$ generator index, the only way this index can be matched in singlet structures such as $\partial \cdot j$ and $G\widetilde{G}$ is through the trace $\mathrm{Tr}(t^a)$. Consequently, we obtain
\begin{equation}
    [\mathcal O^a_E] = C_{EA}  \,[\partial\cdot j^a_A] + C_{EA}^{\tr}  \Tr(t^a) \,[\partial\cdot j_A] + C_{EG} \Tr(t^a) \, [G \widetilde G] + \text{ev}\,,
\end{equation}
where `ev' denotes evanescent operators whose Green's functions with external fields vanish in $d \to 4$. To determine the coefficients, we have to insert these operators in $\psi\bar \psi$ and $GG$ external state matrix elements~\cite{Bos:1992nd, Bhattacharya:2015rsa}, which are denoted as $\expval{ \cdots}_\psi$ and $\expval{ \cdots}_G$, respectively. We will see that these coefficients only depend on gauge couplings, do not depend on the IR regulator or masses, even for the massive case, just like the renormalization constant. Since the physical projections of the evanescent operators always start from higher order corrections, so at the tree level, all the coefficients vanish. Inserting into the one-loop and two-loop matrix elements, we find~\footnote{The bare vector WI is satisfied in $d\neq 4$ keeping the evanescent operator in Eq.~\eqref{eq:bareWardoperatorlevel}. However, it can be ignored in the sense that it has vanishing projections along the physical operators. Thus, no finite renormalization happens in the vector WI. Our computations show this explicitly.}
\begin{equation}\label{eq:ev_projections}
\begin{aligned}
    C_{EA} & =K^{(1)}_\psi  \frac{\alpha_s}{4\pi}  +  \qty(K^{(2)}_\psi - K^{(1)}_\psi F^{(1)}_\psi)\left(\frac{\alpha_s}{4\pi}\right)^2 \,, \\
    C_{EA}^{\tr} & = \qty(K^{\tr (2)}_\psi - K^{(1)}_G A^{(1)}_\psi) \left(\frac{\alpha_s}{4\pi}\right)^2 \,,  \\
    C_{EG} & = K^{(1)}_G \frac{\alpha_s}{4\pi} + \order{\alpha_s^2}\,.
\end{aligned}
\end{equation}
The $K_\psi$, $K^{\tr}_\psi$ and $K_G$ are the projections along the $\overline
{\text{MS}}$ non-single current, singlet current, and anomaly operator of the $\expval {[\mathcal O_E]}_\psi$ and $\expval {[\mathcal O_E]}_G$ matrix elements. Similarly $F_\psi$ and $A_\psi$ are projections along a singlet and non-singlet current of $\expval {[\mathcal \partial \cdot j]}_\psi$ and $\expval {[G \widetilde G]}_\psi$, respectively. Plugging $[\mathcal O^a_E]$ into the WI given in Eq.~\eqref{eq:renorm_WI}, we get
\begin{equation}\label{eq:projected_WI}
    (1 - C_{EA})  \,[\partial\cdot j^a_A] - C_{EA}^{\tr}  \Tr(t^a) \,[\partial\cdot j_A] = [\mathcal O^a_{\text{eom}}] + C_{EG} \Tr(t^a) \, [G \widetilde G] + \text{ev} \,.
\end{equation}

We define the renormalized physical current $j^{aR}_{A\mu}$ and the renormalized anomaly operator $(G \widetilde G)^R$, which are related to the $\overline{\text{MS}}$ operators, as
\begin{equation}\label{eq:finite_renorm}
\begin{aligned}
    j^{aR}_{A\mu} & = z_A  \,[j^a_{A\mu}] + z_A^{\tr}  \Tr(t^a) \,[j_{A\mu}]\,, \quad z_A = (1 - C_{EA})\,, \quad z_A^{\tr} = - C_{EA}^{\tr}\,; \\
    &\quad\quad\quad\quad(G \widetilde G)^R  = z_G[G \widetilde G]\,, \quad  -\frac{\alpha_s}{4\pi} 2T_F \,z_G = C_{EG}\,.
\end{aligned}
\end{equation}
The $z_A$, $z_A^{\tr}$, $z_G$ are called finite renormalization. Each of the currents in a triangle diagram is RG invariant, there is no overall counterterm, the anomaly coefficient must be gauge coupling independent, which implies $z_G$ must always be 1. Substituting this into Eq.~\eqref{eq:projected_WI} gives the Adler-Bell-Jackiw (ABJ) anomaly in $d = 4$
\begin{equation}\label{eq:ABJ_anomaly}
    \partial \cdot j^{aR}_A = [\mathcal O^a_{\text{eom}}] - 2T_F \Tr(t^a) \frac{\alpha_s}{4\pi} (G \widetilde G)^R \,.
\end{equation}

This anomaly equation is true for all orders in $\alpha_s$, which means that the coefficient $G\widetilde G$ does not receive any perturbative correction.

\subsection{Anomalous Dimension Matrix and RG Evolutions}

The renormalized current is  related to the bare current by
\begin{equation}\label{eq:finite_renorm1}
    j^{aR}_{A\mu} = z_A \, \mathcal{Z}_A \, j^{a}_{A\mu}  + (z_A \mathcal{Z}_A^{\tr} + z_A^{\tr}\mathcal{Z}_A^s) \Tr(t^a) j_{A\mu} \,,
\end{equation}
where $\mathcal Z_A$, $\mathcal Z_A^s$ are renormalization constants corresponding to non-singlet and singlet currents, and $\mathcal Z_A^{\tr}$ is a singlet projection to have renormalized the non-singlet current.  On the other hand,
\begin{equation}
    [G \widetilde G]  = \mathcal Z_G \, G\widetilde G + \mathcal Z_{GA} \, \partial \cdot j_{A}\;.
\end{equation}

Taking the derivative with respect to $\mu$ on both sides of Eq.~\eqref{eq:ABJ_anomaly} and using the RG equation $\mu \frac{d}{d\mu} [\mathcal O] =  \hat\gamma [\mathcal O]$, where $\hat \gamma = \mu \dv{\hat{\mathcal Z}}{\mu} \hat {\mathcal Z}^{-1} $, one can obtain
\begin{equation}\label{eq:ADMs_restr}
    \gamma_{A} \, \partial \cdot j^{aR}_{A} + \gamma_{A}^{\tr} \Tr(t^a)\partial \cdot  j^{R}_{A} = - 2T_F \Tr(t^a) \frac{\alpha_s}{4\pi} \qty[\qty(\frac{\beta(\alpha_s)}{\alpha_s} + \gamma_G)[G \widetilde G] + \gamma_{GA} \, \partial \cdot  j^{R}_{A}]\,,
\end{equation}
it can have many solutions; the simplest one is 
\begin{equation}
    \gamma_G = - \frac{\beta(\alpha_s)}{\alpha_s}\,, \quad  \gamma_{A}^{\tr} = - 2T_F \frac{\alpha_s}{4\pi}\gamma_{GA}\,,  \quad \gamma_A = 0\,.
\end{equation}
The first relation is supported by direct results \cite{Larin:1993tq}. We will see below that the last two relations of ADM also hold. Since the anomaly is already of order $\alpha_s$, to find $z_A$ (or $z_A^{\tr}$) up to order $\alpha_s^2$, it is sufficient to calculate the bare Green function corresponding to the anomaly up to one-loop,
\begin{equation}\label{eq:GGdual_green_fn}
    \expval{G \widetilde G}_\psi = \qty[-6 C_F\qty(\frac{1}{\epsilon} + 2 + \log(\frac{\mu^2}{-p^2}))\frac{\alpha_s}{4\pi}] \expval{[\partial \cdot j_A]}_\psi \,,
\end{equation}

where the coefficient of the pole gives the $\gamma_{GA}$, which is written as 
\begin{equation}
    \gamma_{GA} = -  2\epsilon \alpha_s \frac{\partial \mathcal Z_{GA}}{\partial \alpha_s} = 12 C_F \,\frac{\alpha_s}{4\pi} \,.
\end{equation}

Since the {\it Chern-Simons} current ($K_\mu$), defined as $\partial \cdot K = G\widetilde G$, is a gauge-variant operator, it does not mix with the gauge invariant current $j_{A\mu}^a$, but the non-singlet current can mix with a single current; therefore, the renormalization of these currents can be written in a matrix form as
\begin{equation}\label{eq:current_renorm}
    \mqty(j^{aR}_{A\mu} \\ \Tr(t^a) j^R_{A\mu}) = \hat z_A \hat  {\mathcal Z_A} \mqty( j^{a}_{A} \\ \Tr(t^a)  j_{A\mu}), \quad \hat z_A = \mqty(z_A & z_A^{\tr} \\ 0 & z_A^s), \quad \hat {\mathcal Z_A} =  \mqty(\mathcal Z_A & \mathcal Z_A^{\tr} \\ 0 & \mathcal Z_A^s)\,.
\end{equation}

The finite renormalization ($z_A^s$) and renormalization constant ($\mathcal Z_A^s$) for the single current can be found by adding the two contributions of the non-single current and multiplying by the number of generations ($n_g$) factor with the `tr' part, i.e. just by taking $t^a$ to be an identity matrix. The RG evolution of these currents,
\begin{equation}
    \mu\dv{\mu}\mqty(j^{aR}_{A\mu} \\ \Tr(t^a) j^R_{A\mu}) = \hat\gamma_A \mqty(j^{aR}_{A\mu} \\ \Tr(t^a)  j^R_{A\mu}) , \quad  \hat\gamma_A = \mqty(\gamma_A & \gamma_A^{\tr} \\ 0 & \gamma_A^s) \,,
\end{equation}
where the ADM matrix is
\begin{equation}
    \hat\gamma_A =  \qty[\mu \dv{\mu} \qty(\hat z_A \hat  {\mathcal Z_A})] \qty(\hat z_A \hat  {\mathcal Z_A} )^{-1} =  - 2\epsilon\alpha_s  \hat z_A \frac{\partial \hat{\mathcal Z_{A}}}{\partial \alpha_s} \hat z_A^{-1} + 2\beta(\alpha_s) \frac{\partial \hat z_{A}}{\partial \alpha_s} \hat z_A^{-1} \,.
\end{equation}
In order to get the expression for the ADM, we first determine the finite renormalizations defined in Eqs.~\eqref{eq:finite_renorm} and \eqref{eq:ev_projections}, where $K_\psi$ (or $F_\psi$) and $K_\psi^{\tr}$ represent the coefficients of $I \odot t^a$ and $I \Tr(t^a)$ within the $K$ (or $F$) matrix (see Eqs.~\eqref{eq:finite_pieces_1l} and \eqref{eq:finite_pieces_2l}), $A^{(1)}_\psi$ are the finite pieces of Eq.~\eqref{eq:GGdual_green_fn}, and $K_G^{(1)} = - 2T_F$ corresponds to the one-loop anomaly coefficient. An analogous notation is used for the renormalization matrix $\mathcal Z$. Therefore,
\begin{equation}
\begin{aligned}
    \mathcal Z_A & = 1 + 2C_F\beta_0  \frac{\alpha_s^2}{16\pi^2}\frac{1}{\epsilon}\,, \qquad \mathcal Z_A^{\tr} = 6C_F T_F\frac{\alpha_s^2}{16\pi^2}\frac{1}{\epsilon}\,, \qquad \mathcal Z_A^s = \mathcal Z_A + n_g \mathcal Z_A^{\tr}\,,\\
    z_A & = 1 - 4C_F \frac{\alpha_s}{4\pi} -  \frac{C_F}{9} \qty(107 C_A - 198C_F - 4 n_F T_F) \frac{\alpha_s^2}{16\pi^2}\,, \\ z_A^{\tr} & = - 3C_F T_F \frac{\alpha_s^2}{16\pi^2}\,,  \qquad z_A^s = z_A + n_g z_A^{\tr}\,.
\end{aligned}
\end{equation}

Finally, we obtain the ADMs for the current 
\begin{equation}
\begin{aligned}
    \gamma_{A} &  = - 2\epsilon\alpha_s  \frac{\partial \mathcal Z_{A}}{\partial \alpha_s} + 2\beta(\alpha_s) z_A^{-1} \frac{\partial z_{A}}{\partial \alpha_s} = 0\,, \\
    \gamma_{A}^{\tr} & = - 2\epsilon\alpha_s \frac{\partial \mathcal Z_{A}^{\tr}}{\partial \alpha_s} + 2\beta(\alpha_s) \frac{1}{z_A z_A^s} \qty(z_A\frac{\partial z_{A}^{\tr}}{\partial \alpha_s} - z_A^{\tr}\frac{\partial z_{A}}{\partial \alpha_s}) = - 24 C_F T_F \, \frac{\alpha_s^2}{16\pi^2}\,, \\
    \gamma_A^s & = \gamma_{A} + n_g\gamma_{A}^{\tr} \,.
\end{aligned}
\end{equation}
This result exactly matches the expectation of Eq.~\eqref{eq:ADMs_restr}. 
By imposing the scale invariance condition on the Lagrangian Eq.~\eqref{eq:lag_mass_basis}, 
\begin{equation}
    \mu \dv{\mu} (- c^b_A\,a\,\partial \cdot j^{bR}_{A} + \mathcal C_G\,a\, (G\widetilde G)^R) = 0\,, \qquad c_A^b = c_R^b - c_L^b \,,
\end{equation}
where scale dependency of the axion field is ignored since it starts from an additional $\order {1/f_a}$ correction, alongside using the RG equations for the renormalized operators, and accounting for both the up and down quark sector axial currents, we can easily obtain the running of the Wilson coefficients,
\begin{equation}
\begin{aligned}
    \mu\dv{c^a_{uA}}{\mu} \, t^a & = - \gamma_A \, c^a_{uA} \, t^a - \gamma_A^{\tr}\qty(c_{uA}^a + c_{dA}^a)\Tr(t^a) I + \gamma_{GA}\, \mathcal C_G\, I \,, \\
    \mu\dv{c^a_{dA}}{\mu} \, t^a & = - \gamma_A \, c^a_{dA} \, t^a - \gamma_A^{\tr}\qty(c_{uA}^a + c_{dA}^a)\Tr(t^a) I + \gamma_{GA}\, \mathcal C_G\, I \,, \\
    \mu\dv{\mathcal C_G}{\mu} & =  -\gamma_G \,\mathcal C_G\,, \qquad  c_{uA}^a \, t^a = \mathcal C_{uA}\,, \quad c_{dA}^a \, t^a = \mathcal C_{dA}\,.
\end{aligned}
\end{equation}

The trace component includes both types of Wilson coefficients because the up- and down-type non-singlet currents mix with both types of singlet currents via triangle diagrams. Here, $\mathcal C_G$, $\mathcal C_{uA}$ and $\mathcal C_{dA}$ are the same as in our previous paper's notation $\mathcal C_1$, $\mathcal C_{2A}$ and $\mathcal C_{3A}$, respectively \cite{Bisht:2024hbs}. Some important checks of our result are required.

\subsection{Consistency Checks}
Several checks validate our calculations. Most importantly, the vector and axial bare Ward identities have been explicitly verified for both the pole and finite terms up to two loops in the BMHV scheme. This is the most robust and non-trivial check. Some additional smaller checks are listed below
\begin{enumerate}
\item Quark wavefunction renormalization does not involve a $\gamma_5$ and must be the same in NDR and BMHV. Indeed, we found this explicitly to be true up to $\mathcal{O}(\alpha_s^2)$ (table~\ref{tab:SE2loop}).

\item By the vector Ward identity, the poles of current insertion must be canceled by wavefunction renormalization. This has been explicitly verified at one and two loops.

\item In the $\overline{\text{MS}}$ scheme, the operator renormalization constants satisfy the relation~\cite{Buras:2020xsm}
\begin{equation}
    4\mathcal{Z}^{(2)}_{\epsilon^2} + 2\beta_0 \mathcal{Z}^{(1)}_{\epsilon} - 2\mathcal{Z}^{(1)}_{\epsilon}\mathcal{Z}^{(1)}_{\epsilon} = 0\;,
\end{equation}
where the subscript denotes the pole order. In particular, this relation ensures that the double pole $1/\epsilon^2$ at the two-loop level must vanish if the corresponding single pole at one-loop vanishes and there is no contribution from the last mixing term. This is true for both vector and axial-vector currents in our analysis, as demonstrated by the entries of table~\ref{tab:SE2loop} and ~\ref{tab:J2loop}.

\item No $1/\epsilon^2$ poles appear for physical projections of the evanescent operator (table~\ref{tab:EV2loop}), as the Dirac algebra with an evanescent insertion must necessarily give a $d-4$ proportional factor when projected onto a non-evanescent Dirac structure.

\item The finite renormalization of the axial current is found to be in agreement with the original Larin result~\cite{Larin:1993tq} up to $\mathcal{O}(\alpha_s^2)$.

\end{enumerate}

\section{Conclusion}\label{sec:conclusion}
In this work, we have developed a systematic framework for the renormalization of effective field theories of axions within the BMHV scheme, with particular emphasis on the role of evanescent operators and their interplay with Ward identities. Starting from the observation that fermionic dimension-five operators in axion EFT can be expressed in terms of Standard Model flavor currents, we have demonstrated that the renormalization of these operators is governed entirely by the renormalization properties of the underlying chiral currents. This provides a conceptually clean route to studying higher-order effects in axion phenomenology.

A central result of our analysis is the explicit derivation of bare and renormalized Ward identities in the BMHV scheme. We showed that, away from four dimensions, the naive Ward identities are violated due to the presence of evanescent operators originating from the non-anti-commuting nature of $\gamma_5$. These operators arise already at the level of current divergences and must be consistently included in the operator basis. Importantly, we established that the equation-of-motion operators remain finite and do not mix under renormalization, while the evanescent operators induce non-trivial mixing into physical operators. This structure provides a transparent understanding of how BMHV restores consistency at the quantum level.

We have verified the validity of the bare Ward identities explicitly up to the two-loop order $\mathcal{O}(\alpha_s^2)$ by computing matrix elements with external fermions. The agreement between the left- and right-hand sides of the Ward identities, including both pole and finite parts, serves as a highly non-trivial check of our framework. Furthermore, by projecting the evanescent operators onto physical operators, we have derived the finite renormalization constants required to restore the four-dimensional renormalized Ward identities. In particular, we recover the well-known structure of axial current renormalization and the emergence of the ABJ anomaly in a fully consistent BMHV setup.

An important conceptual outcome of this work is the clarification of the role of evanescent operators in chiral theories. Although they vanish in four dimensions, their effects persist in intermediate steps of dimensional regularization and influence physical quantities through operator mixing. This is especially relevant in effective field theories, where such operators contribute to anomalous dimensions starting at higher-loop order. Our analysis makes this mechanism explicit in the context of axion EFT and connects it directly to the structure of Ward identities.

From a phenomenological perspective, our results provide a robust foundation for precision studies of the interactions of axion- and axion-like particles. The consistent treatment of $\gamma_5$ and the systematic inclusion of evanescent operators are essential for reliable multi-loop computations, particularly when matching high-scale theories onto low-energy effective descriptions. Our framework can be readily extended to include electroweak corrections and other operator classes, which will be important for future applications.

In summary, we have presented a coherent and self-consistent implementation of the BMHV scheme in axion effective field theory, highlighting the essential role of evanescent operators in maintaining the consistency of Ward identities and renormalization. This work bridges formal aspects of dimensional regularization with practical EFT computations and lays the groundwork for further developments in precision axion physics.

\acknowledgments
The authors thank Joydeep Chakrabortty, Sanmay Ganguly, and Nilay Kundu for helpful discussions. The authors also thank Goutam Das for comments on the draft. SC thanks the Science and Engineering Research Board, Government of India (Grant No. SRG/2023/001162), and the IIT-Kanpur initiation grant (PHY/2022220) for financial support. AS acknowledges the hospitality of ICTS, Bengaluru, during the ``School for Advanced Topics in Particle Physics (SATPP): QCD, Effective Field Theories, and Nuclear Physics", where the final part of this work was completed.

\appendix

\section{BMHV Scheme}\label{app:HValgebra}
In this appendix, we collect standard properties and identities in the BMHV scheme. Some other recent references include~\cite{Belusca-Maito:2023wah, Naterop:2023dek}. 
\subsection{Algebra and Identities}
Decomposition in $4$ and $d-4$ dimensions:
\begin{equation}
    \gamma^\mu = \widetilde{\gamma}^\mu +  \widehat{\gamma}^\mu \;, \quad g^{\mu\nu} = \widetilde{g}^{\mu\nu} + \widehat{g}^{\mu\nu} \;, \quad p^\mu = \widetilde{p}^\mu + \widehat{p}^\mu\;.
\end{equation}
(Anti)commutation identities
\begin{equation}
\begin{alignedat}{2}
    \{\gamma^\mu, \gamma^\nu \} &= 2g^{\mu\nu}\;, &\quad \{\widetilde{\gamma}^\mu, \widetilde{\gamma}^\nu \} &= \{\gamma^\mu, \widetilde{\gamma}^\nu \} = 2\widetilde{g}^{\mu\nu}\;,  \\
    \{\widetilde{\gamma}^\mu, \widehat{\gamma}^\nu \} &= 0 \;, &\quad
    \{\widehat{\gamma}^\mu, \widehat{\gamma}^\nu \} &= \{\gamma^\mu, \widehat{\gamma}^\nu \} = 2\widehat{g}^{\mu\nu} 
    \;, \\ 
    \{\widetilde{\gamma}^\mu, \gamma_5 \} &= [\widehat{\gamma}^\mu, \gamma_5] = 0\;, &\quad
    \{\gamma^\mu, \gamma_5 \} &= \{\widehat{\gamma}^\mu, \gamma_5 \} = 2\widehat{\gamma}^\mu \gamma_5 \;.
\end{alignedat}
\end{equation}
Contractions of metric with Dirac matrices:
\begin{equation}
    \begin{aligned}
        g_{\mu\nu}\gamma^\nu &= \gamma_\mu\;, \\
        \widetilde{g}_{\mu\nu}\widetilde{\gamma}^\nu &= g_{\mu\nu} \widetilde{\gamma}^\nu = \widetilde{g}_{\mu\nu}\gamma^\nu = \widetilde{\gamma}_\mu\;, \\
        \widehat{g}_{\mu\nu}\widehat{\gamma}^\nu &= g_{\mu\nu} \widehat{\gamma}^\nu = \widehat{g}_{\mu\nu}\gamma^\nu = \widehat{\gamma}_\mu\;,\\
        \widetilde{g}_{\mu\nu}\widehat{\gamma}^\nu &= \widehat{g}_{\mu\nu}\widetilde{\gamma}^\nu = 0\;.
    \end{aligned}
\end{equation}
The same exists for four momenta as well i.e. replace $\gamma$ everywhere by $p$:
\begin{equation}
\gamma^\nu \to p^\nu\;, \widehat{\gamma}^\nu \to \widehat{p}^\nu\;, \widetilde{\gamma}^\nu \to \widetilde{p}^\nu\;.
\end{equation}
Contractions of metric with itself and of Dirac matrices with themselves
\begin{equation}
    \begin{aligned}
        g^{\mu\nu}g_{\mu\nu} &= d\;, \qquad &  \gamma^\mu \gamma_\mu &= d\;,\\
        \widetilde{g}^{\mu\nu} \widetilde{g}_{\mu\nu} &= g^{\mu\nu} \widetilde{g}_{\mu\nu} = \widetilde{g}^{\mu\nu} g_{\mu\nu} = 4\;,  \qquad &  \widetilde{\gamma}^\mu \widetilde{\gamma}_\mu &= \gamma^\mu \widetilde{\gamma}_\mu = \widetilde{\gamma}^\mu \gamma_\mu = 4\;,\\
        \widehat{g}^{\mu\nu} \widehat{g}_{\mu\nu} &= g^{\mu\nu} \widehat{g}_{\mu\nu} = \widehat{g}^{\mu\nu} g_{\mu\nu} = d-4\;,  \qquad &  \widehat{\gamma}^\mu \widehat{\gamma}_\mu &= \gamma^\mu \widehat{\gamma}_\mu = \widehat{\gamma}^\mu \gamma_\mu = d-4\;,\\
        \widetilde{g}^{\mu\nu}\widehat{g}_{\mu\nu} &= \widehat{g}^{\mu\nu}\widetilde{g}_{\mu\nu} = 0\;, \qquad & \widetilde{\gamma}^\mu \widehat{\gamma}_\mu &= \widehat{\gamma}^\mu \widetilde{\gamma}_\mu = 0\;.
    \end{aligned}
\end{equation}
Furthermore, the following contractions are useful
\begin{equation}
    \begin{alignedat}{2}
        \widehat{\gamma}^\mu \widehat{\gamma}^\nu \widehat{\gamma}_\mu &= (6-d)\widehat{\gamma}^\nu \;, &\quad \widehat{\gamma}^\mu \widetilde{\gamma}^\nu \widehat{\gamma}_\mu &= (4-d)\widetilde{\gamma}^\nu \;,\\
        \widetilde{\gamma}^\mu \widetilde{\gamma}^\nu \widetilde{\gamma}_\mu &= -2\widetilde{\gamma}^\nu \;, &\quad  \widetilde{\gamma}^\mu \widehat{\gamma}^\nu \widetilde{\gamma}_\mu &= -4\widehat{\gamma}^\nu \;.
    \end{alignedat}
\end{equation}

\subsection{Conventions and Identities Involving \texorpdfstring{$\epsilon^{\mu\nu\rho\sigma}$}{epsilon} and \texorpdfstring{$\gamma_5$}{gamma5}}\label{app:conventions}
Following~\cite{Bohm:2001yx},  for the Levi-Civita tensor $\epsilon^{\mu\nu\rho\sigma}$, we use the convention $\epsilon^{0123} = - \epsilon_{0123}= +1$. This is consistent with the following definition of $\gamma_5$ and the four-dimensional trace

\begin{equation}\label{eq:defg5}
            \gamma_5 \equiv -\frac{i}{4!} \epsilon_{\mu\nu\rho \sigma} \widetilde{\gamma}^\mu \widetilde{\gamma}^\nu \widetilde{\gamma}^\rho \widetilde{\gamma}^\sigma \;,\quad   \text{Tr}(\widetilde{\gamma}^\mu \widetilde{\gamma}^\nu \widetilde{\gamma}^\rho \widetilde{\gamma}^\sigma \gamma_5) = -4i\epsilon^{\mu\nu\rho\sigma} \;.
\end{equation}
Another useful identity is
\begin{equation}
    \frac{1}{2}( \gamma_\mu \gamma_5 - \gamma_5 \gamma_\mu ) = -\frac{i}{3!} \epsilon_{\mu \nu\rho\sigma} \gamma^\nu \gamma^\rho \gamma^\sigma\;.
\end{equation}

\section{Feynman Diagrams and Computational Details}\label{app:diagrams}
In this Appendix, we present important details of our diagrammatic calculations. The existing literature on axial current renormalization with non-anticommuting $\gamma_5$~\cite{Chen:2022lun, Ahmed:2021spj, Larin:1991tj, Larin:1993tq} exclusively follows Larin's prescription. Here, the matrix elements with operator insertions are projected onto the physical currents by the trace operation to simplify calculations. We use an alternative procedure and directly evaluate the diagrams for the matrix elements using the BMHV scheme algebra in the same spirit as ~\cite{Collins:1984xc, Bos:1992nd}. We present the relevant diagrams and the pole coefficients in dimensional regularization ($d=4-2\epsilon$). 

\subsection{One-loop \texorpdfstring{$\mathcal{O}(\alpha_s)$}{O[alpha s]}}
The one-loop calculations for the current, self-energy, and evanescent operators (in BMHV) are standard and straightforward. Only one diagram exists for the current insertion. The kinetic part of the evanescent operator is also inserted in the same diagram. The self-energy terms also have only one diagram, and the gluon-fermion term of the evanescent operator is inserted in the respective vertices. The corresponding poles are given in Eq.~\eqref{eq:poles_1l} and the finite pieces in Eq.~\eqref{eq:finite_pieces_1l}. 
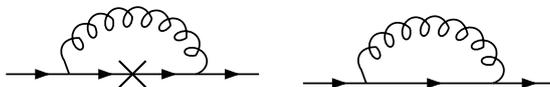
\begin{figure}[H]
\centering
\begin{subfigure}{0.25\textwidth}
\centering
\resizebox{1.0\textwidth}{!}{
\begin{tikzpicture}
    \begin{feynman}
    \vertex (a);
    \vertex [right=0.5cm of a] (b) ;
    \vertex [right=0.5cm of b] (c);
    \vertex [right=0.5cm of c] (d) ;
    \vertex [right=0.5cm of d] (e) ;
    \vertex [below = 0.5cm of c] (f) ;
    \diagram* {
        (a) -- [fermion] (b) -- [fermion] (c) -- [fermion, insertion = 0] (d) -- [fermion] (e), (d) -- [gluon, half right] (b)
    };
    \end{feynman}
\end{tikzpicture}
}
\end{subfigure}
\begin{subfigure}{0.25\textwidth}
\centering
\resizebox{1.0\textwidth}{!}{
\begin{tikzpicture}
    \begin{feynman}
    \vertex (a);
    \vertex [right=0.5cm of a] (b) ;
    \vertex [right=0.5cm of b] (c);
    \vertex [right=0.5cm of c] (d) ;
    \vertex [right=0.5cm of d] (e) ;
    \diagram* {
        (a) -- [fermion] (b) -- [fermion] (d) -- [fermion] (e), (d) -- [gluon, half right] (b)
    };
    \end{feynman}
\end{tikzpicture}
}
\end{subfigure}
\caption{Diagrams at one-loop, cross represents current insertion.}
\label{fig:oneloopdia}
\end{figure}

\subsection{Two-loop \texorpdfstring{$\mathcal{O}(\alpha_s^2)$}{O[alpha  s squared]}}
In this section, we present the two-loop calculations which have been automated using the following computational workflow. We use \texttt{FeynRules}~\cite{Christensen:2008py} to produce Feynman rules in \texttt{UFO}~\cite{Degrande:2011ua} format for Lagrangian and field definitions. The two-loop and counterterm diagrams are generated by \texttt{QGRAF}~\cite{Nogueira:1991ex} and the corresponding amplitudes in the Lorentz and color parts are generated using \texttt{tapir}~\cite{Gerlach:2022qnc}. The subsequent handling of the Dirac algebra, the color algebra and the tensor reduction is done using \texttt{FORM}~\cite{Vermaseren:2000nd, Tentyukov:2007mu}. A 6-rank tensor reduction was needed, which required solving 76 linear equations and was performed using \texttt{Ginac}~\cite{Vollinga:2005pk}.  Finally, we use \texttt{Kira}~\cite{Lange:2025fba} and \texttt{Reduze}~\cite{Studerus:2009ye} for IBP reduction to the basis of two-loop master integrals.

\begin{figure}[t]
\centering
\begin{subfigure}{0.2\textwidth}
\centering
\resizebox{1.0\textwidth}{!}{
\begin{tikzpicture}
    \begin{feynman}
    \vertex (a) ;
    \vertex [right=0.5cm of a] (b);
    \vertex [right=1cm of b] (c);
    \vertex [right=1cm of c] (d);
    \vertex [right=0.5cm of d] (e);
    \vertex [above=0.7cm of c] (f) ;
    \vertex [above=0.5cm of f] (g) ;
     \vertex [below=0.5cm of c] (h) ;
    \diagram* {
            (a) -- [fermion] (b) -- [fermion] (c) -- [fermion, insertion = 0] (d) -- [fermion] (e), (f) -- [gluon, quarter right] (b), (d) -- [gluon, quarter right] (f), (f) -- [gluon, half right] (g), (g) -- [gluon, half right] (f)
        };
        \end{feynman}
\end{tikzpicture}
}
\caption{J1}
\end{subfigure}
\hfill
\begin{subfigure}{0.22\textwidth}
\centering
\resizebox{1.0\textwidth}{!}{
\begin{tikzpicture}
    \begin{feynman}
    \vertex (a) ;
    \vertex [right=0.5cm of a] (b);
    \vertex [right=0.7cm of b] (c);
    \vertex [right=0.7cm of c] (i);
    \vertex [right=0.8cm of i] (d);
    \vertex [right=0.7cm of d] (e);
    \vertex [right=0.5cm of e] (f) ;
    \vertex [below=0.8cm of i] (h)  ;
    \diagram* {
            (a) -- [fermion] (b) -- [fermion] (c) -- [fermion] (i) -- [fermion, insertion = 0] (d) -- [fermion] (e) -- [fermion] (f), (d) -- [gluon, half right] (c), (e) -- [gluon, half right] (b)
        };
        \end{feynman}
\end{tikzpicture}
}
\caption{J2[3]}
\end{subfigure}\hfill
\begin{subfigure}{0.24\textwidth}
\centering
\resizebox{1.0\textwidth}{!}{
\begin{tikzpicture}
    \begin{feynman}
    \vertex (a)  ;
    \vertex [right=0.7cm of a] (b);
    \vertex [right=0.7cm of b] (c);
    \vertex [right=0.8cm of c] (d);
    \vertex [right=0.7cm of d] (i);
    \vertex [right=0.7cm of i] (e);
    \vertex [right=0.7cm of e] (f)  ;
    \vertex [below=0.8cm of i] (h)  ;
    \diagram* {
            (a) -- [fermion] (b) -- [fermion] (c) -- [fermion] (d) -- [fermion] (i) -- [fermion, insertion = 0] (e) -- [fermion] (f), (b) -- [gluon, half right] (d), (e) -- [gluon, half right] (c)
        };
        \end{feynman}
\end{tikzpicture}
}
\caption{J3[3]}
\end{subfigure}
\hfill
\begin{subfigure}{0.24\textwidth}
\centering
\resizebox{1.0\textwidth}{!}{
\begin{tikzpicture}
    \begin{feynman}
    \vertex (a)  ;
    \vertex [right=0.7cm of a] (b);
    \vertex [right=0.7cm of b] (c);
    \vertex [right=0.8cm of c] (d);
    \vertex [right=0.7cm of d] (i);
    \vertex [right=0.7cm of i] (e);
    \vertex [right=0.7cm of e] (f)  ;
    \vertex [below=0.8cm of i] (h)  ;
    \diagram* {
            (a) -- [fermion] (b) -- [gluon] (c) -- [gluon] (d) -- [fermion] (i) -- [fermion, insertion = 0] (e) -- [fermion] (f), (b) -- [fermion, half right] (d), (e) -- [gluon, half right] (c)
        };
        \end{feynman}
\end{tikzpicture}
}
\caption{J4[2]}
\end{subfigure}

\begin{subfigure}{0.22\textwidth}
\centering
\resizebox{1.0\textwidth}{!}{
\begin{tikzpicture}
    \begin{feynman}
    \vertex (a)  ;
    \vertex [right=0.5cm of a] (b);
    \vertex [right=1cm of b] (c);
    \vertex [right=1cm of c] (d);
    \vertex [right=0.5cm of d] (e) ;
    \vertex [above right=1cm of b] (f) ;
    \vertex [above left=1cm of d] (g) ;
     \vertex [below=0.5cm of c] (h) ;
    \diagram* {
            (a) -- [fermion] (b) -- [fermion] (c) -- [fermion, , insertion = 0] (d) -- [fermion] (e), (f) -- [gluon, quarter right] (b), (d) -- [gluon, quarter right] (g), (f) -- [gluon, half right] (g),(g) -- [gluon, half right] (f)
        };
        \end{feynman}
\end{tikzpicture}
}
\caption{J5}
\end{subfigure}
\begin{subfigure}{0.22\textwidth}
\centering
\resizebox{1.0\textwidth}{!}{
\begin{tikzpicture}
    \begin{feynman}
    \vertex (a)  ;
    \vertex [right=0.5cm of a] (b);
    \vertex [right=1cm of b] (c);
    \vertex [right=1cm of c] (d);
    \vertex [right=0.5cm of d] (e) ;
    \vertex [above right=1cm of b] (f) ;
    \vertex [above left=1cm of d] (g) ;
     \vertex [below=0.5cm of c] (h) ;
    \diagram* {
            (a) -- [fermion] (b) -- [fermion] (c) -- [fermion, , insertion = 0] (d) -- [fermion] (e), (f) -- [gluon, quarter right] (b), (d) -- [gluon, quarter right] (g), (f) -- [ghost, half right] (g),(g) -- [ghost, half right] (f)
        };
        \end{feynman}
\end{tikzpicture}
}
\caption{J6}
\end{subfigure}
\begin{subfigure}{0.22\textwidth}
\centering
\resizebox{1.0\textwidth}{!}{
\begin{tikzpicture}
    \begin{feynman}
    \vertex (a)  ;
    \vertex [right=0.5cm of a] (b);
    \vertex [right=1cm of b] (c);
    \vertex [right=1cm of c] (d);
    \vertex [right=0.5cm of d] (e) ;
    \vertex [above right=1cm of b] (f) ;
    \vertex [above left=1cm of d] (g) ;
     \vertex [below=0.5cm of c] (h) ;
    \diagram* {
            (a) -- [fermion] (b) -- [fermion] (c) -- [fermion, , insertion = 0] (d) -- [fermion] (e), (f) -- [gluon, quarter right] (b), (d) -- [gluon, quarter right] (g), (f) -- [fermion, half right] (g),(g) -- [fermion, half right] (f)
        };
        \end{feynman}
\end{tikzpicture}
}
\caption{J7[6]}
\end{subfigure}\\
\begin{subfigure}{0.22\textwidth}
\centering
\resizebox{0.9\textwidth}{!}{
\begin{tikzpicture}
    \begin{feynman}
    \vertex (a)  ;
    \vertex [right=0.8cm of a] (b) ;
    \vertex [above right=0.8cm of b] (c) ;
    \vertex [below right=0.8cm of b] (d) ;
    \vertex [right=1cm of c] (e) ;
    \vertex [right=1cm of d] (f) ;
    \vertex [right=0.5cm of e] (g) ;
    \vertex [right=0.5cm of f] (h) ;
    \diagram*{
        (b) -- [fermion](c) -- [fermion] (d) -- [fermion] (b); (e) -- [gluon](c) ; (d) -- [gluon] (f) ; (e) -- [fermion] (f) ; (g) -- [fermion] (e) ; (f) -- [fermion] (h) ;
    };
    \draw[shift={(b)}, thin] (65:0.15) -- (245:0.15);
    \draw[shift={(b)}, thin] (115:0.15) -- (295:0.15);
    \end{feynman}
\end{tikzpicture}
}
\caption{J8}
\end{subfigure}
\begin{subfigure}{0.22\textwidth}
\centering
\resizebox{0.9\textwidth}{!}{
\begin{tikzpicture}
    \begin{feynman}
    \vertex (a)  ;
    \vertex [right=0.8cm of a] (b) ;
    \vertex [above right=0.8cm of b] (c) ;
    \vertex [below right=0.8cm of b] (d) ;
    \vertex [right=1cm of c] (e) ;
    \vertex [right=1cm of d] (f) ;
    \vertex [right=0.5cm of e] (g) ;
    \vertex [right=0.5cm of f] (h) ;
    \diagram*{
        (b) -- [fermion](c) -- [fermion] (d) -- [fermion] (b); (e) -- [gluon](d) ; (c) -- [gluon] (f) ; (e) -- [fermion] (f) ; (g) -- [fermion] (e) ; (f) -- [fermion] (h) ;
    };
    \draw[shift={(b)}, thin] (65:0.15) -- (245:0.15);
    \draw[shift={(b)}, thin] (115:0.15) -- (295:0.15);
    \end{feynman}
\end{tikzpicture}
}
\caption{J9}
\end{subfigure} 
\caption{Crosses indicate current insertions; bracketed numbers show the total possible diagrams for that topology (e.g., J2 represents one of three).}
\label{fig:2loopJdia}
\end{figure}
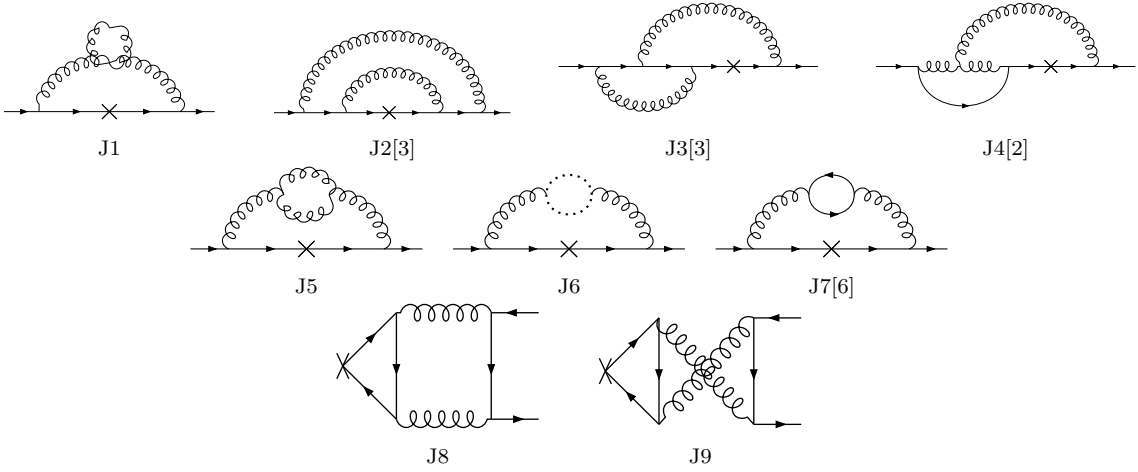

The relevant two-loop diagrams are shown in Fig.~\ref{fig:2loopSEdia}, \ref{fig:2loopJdia}. The corresponding one-loop counterdiagrams are not shown explicitly. For operator insertion Fig.~\ref{fig:2loopJdia}, we show the number of diagrams with the same topology in square brackets. 
\begin{figure}[H]
\centering
\begin{subfigure}{0.24\textwidth}
\centering
\resizebox{1.0\textwidth}{!}{
\begin{tikzpicture}
    \begin{feynman}
    \vertex (a)  ;
    \vertex [right=0.7cm of a] (b);
    \vertex [right=2cm of b] (d);
    \vertex [right=0.7cm of d] (e) ;
    \vertex [above right=1.4cm of b] (f) ;
    \vertex [above=0.5cm of f] (g) ;
    \diagram* {
            (a) -- [fermion] (b) -- [fermion] (d) -- [fermion] (e), (f) -- [gluon, quarter right] (b), (d) -- [gluon, quarter right] (f), (f) -- [gluon, half right] (g), (g) -- [gluon, half right] (f)
        };
        \end{feynman}
\end{tikzpicture}
}
\caption{SE1}
\end{subfigure}
\hfill
\begin{subfigure}{0.24\textwidth}
\centering
\resizebox{1.0\textwidth}{!}{
\begin{tikzpicture}
    \begin{feynman}
    \vertex (a)  ;
    \vertex [right=0.7cm of a] (b);
    \vertex [right=0.7cm of b] (c);
    \vertex [right=0.8cm of c] (d);
    \vertex [right=0.7cm of d] (e);
    \vertex [right=0.7cm of e] (f)  ;
    \diagram* {
            (a) -- [fermion] (b) -- [fermion] (c) -- [fermion] (d) -- [fermion] (e) -- [fermion] (f), (d) -- [gluon, half right] (c), (e) -- [gluon, half right] (b)
        };
        \end{feynman}
\end{tikzpicture}
}
\caption{SE2}
\end{subfigure}
\hfill
\begin{subfigure}{0.24\textwidth}
\centering
\resizebox{1.0\textwidth}{!}{
\begin{tikzpicture}
    \begin{feynman}
    \vertex (a)  ;
    \vertex [right=0.7cm of a] (b);
    \vertex [right=0.7cm of b] (c);
    \vertex [right=0.8cm of c] (d);
    \vertex [right=0.7cm of d] (e);
    \vertex [right=0.7cm of e] (f)  ;
    \diagram* {
            (a) -- [fermion] (b) -- [fermion] (c) -- [fermion] (d) -- [fermion] (e) -- [fermion] (f), (d) -- [gluon, half right] (b), (c) -- [gluon, half right] (e)
        };
        \end{feynman}
\end{tikzpicture}
}
\caption{SE3}
\end{subfigure}
\hfill
\begin{subfigure}{0.24\textwidth}
\centering
\resizebox{1.0\textwidth}{!}{
\begin{tikzpicture}
    \begin{feynman}
    \vertex (a)  ;
    \vertex [right=0.7cm of a] (b);
    \vertex [right=0.7cm of b] (c);
    \vertex [right=0.8cm of c] (d);
    \vertex [right=0.7cm of d] (e);
    \vertex [right=0.7cm of e] (f)  ;
    \diagram* {
            (a) -- [fermion] (b) -- [fermion] (c) -- [gluon] (d) -- [gluon] (e) -- [fermion] (f), (d) -- [gluon, half right] (b), (c) -- [fermion, half right] (e)
        };
        \end{feynman}
\end{tikzpicture}
}
\caption{SE4}
\end{subfigure}
\\
\begin{subfigure}{0.28\textwidth}
\centering
\resizebox{1.0\textwidth}{!}{
\begin{tikzpicture}
    \begin{feynman}
    \vertex (a)  ;
    \vertex [right=0.7cm of a] (b);
    \vertex [right=2cm of b] (d);
    \vertex [right=0.7cm of d] (e) ;
    \vertex [above right=1cm of b] (f) ;
    \vertex [above left=1cm of d] (g) ;
    \diagram* {
            (a) -- [fermion] (b) -- [fermion] (d) -- [fermion] (e), (f) -- [gluon, quarter right] (b), (d) -- [gluon, quarter right] (g), (f) -- [gluon, half right] (g),(g) -- [gluon, half right] (f)
        };
        \end{feynman}
\end{tikzpicture}
}
\caption{SE5}
\end{subfigure}
\hspace{0.2cm}
\begin{subfigure}{0.28\textwidth}
\centering
\resizebox{1.0\textwidth}{!}{
\begin{tikzpicture}
    \begin{feynman}
    \vertex (a)  ;
    \vertex [right=0.7cm of a] (b);
    \vertex [right=2cm of b] (d);
    \vertex [right=0.7cm of d] (e) ;
    \vertex [above right=1cm of b] (f) ;
    \vertex [above left=1cm of d] (g) ;
    \diagram* {
            (a) -- [fermion] (b) -- [fermion] (d) -- [fermion] (e), (f) -- [gluon, quarter right] (b), (d) -- [gluon, quarter right] (g), (f) -- [ghost, half right] (g),(g) -- [ghost, half right] (f)
        };
        \end{feynman}
\end{tikzpicture}
}
\caption{SE6}
\end{subfigure}
\hspace{0.2cm}
\begin{subfigure}{0.28\textwidth}
\centering
\resizebox{1.0\textwidth}{!}{
\begin{tikzpicture}
    \begin{feynman}
    \vertex (a)  ;
    \vertex [right=0.7cm of a] (b);
    \vertex [right=2cm of b] (d);
    \vertex [right=0.7cm of d] (e) ;
    \vertex [above right=1cm of b] (f) ;
    \vertex [above left=1cm of d] (g) ;
    \diagram* {
            (a) -- [fermion] (b)  -- [fermion] (d) -- [fermion] (e), (f) -- [gluon, quarter right] (b), (d) -- [gluon, quarter right] (g), (f) -- [fermion, half right] (g),(g) -- [fermion, half right] (f)
        };
        \end{feynman}
\end{tikzpicture}
}
\caption{SE7}
\end{subfigure}
\caption{Fermion self-energy diagrams at $\mathcal{O}(\alpha_s^2)$}
\label{fig:2loopSEdia}
\end{figure}
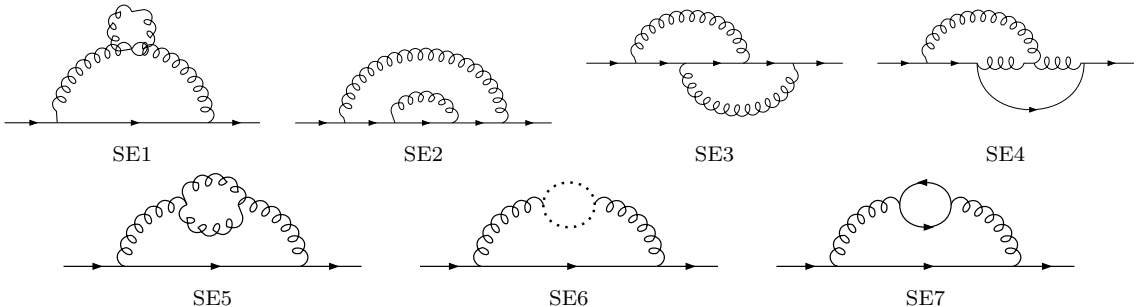
\begin{table}[H]
\centering
\renewcommand{\arraystretch}{1.3}
\begin{tabular}{|l|c|c|c|c|c|c|}
\toprule
\textbf{Diagram} & \textbf{Color factor} & $A$ & $B$ & $\tilde{A}$   & $\tilde{B}$   \\
\midrule
SE1 & * & 0 & 0  & 0  & 0  \\\hline
SE2 & $C_F^2$  & $-1/2$ & $-5/4$  & $1$ & $1$   \\\hline
SE3 & $C_F^2 - C_A C_F/2$ & $1$ & $3/2$ & $2$ & $-2$ \\
SE4 & $C_AC_F$ & $3/2$ & $23/4$  & $-3$  & $-3$ \\\hline
SE5 & $C_A C_F$  & $-1/8$ & $13/16$   &  & \\
SE6 & $C_AC_F$ & $1/8$ & $7/16$ & & \\
SE7 & $C_FT_F$ & $0$ & $- n_F$ &  &  \\
\bottomrule
\end{tabular}
\caption{Pole coefficients of two-loop $\mathcal{O}(\alpha_s^2)$ diagrams. The entries are to be interpreted as $A/\epsilon^2 + B/\epsilon + (2A/\epsilon) \log(-\mu^2/p^2)$ for the two-loop diagram and $\tilde{A}/\epsilon^2 + \tilde{B}/\epsilon + (\tilde{A}/\epsilon) \log(-\mu^2/p^2)$ for the corresponding counterdiagram. An overall coupling factor of $\alpha_s^2/(16\pi^2)$ has been extracted, where $\alpha_s = g_s/(4\pi)$. See the text for the interpretation of the empty entries.}\label{tab:SE2loop}
\end{table}
Tables~\ref{tab:SE2loop}, \ref{tab:J2loop} and \ref{tab:EV2loop} give the color factors and pole coefficients of the two-loop diagrams and counterdiagrams. In Table~\ref{tab:SE2loop}, diagrams SE5 to SE7 have a single counterdiagram, obtained by inserting the pole of gluon wavefunction renormalization (including gluon, ghost, and all quark contributions) into the remaining one-loop diagram. This pole has the structure $\sim p^2g_{\mu\nu} - p_\mu p_\nu $, which, when inserted into the remaining one-loop diagram, gives zero identically. Similarly, in table~\ref{tab:J2loop}, the empty block of counterdiagram entries implies that J5 to J7 have a single counterdiagram, which is identically zero. Entries with hyphen `-' in the counterdiagram coefficient columns indicate that no counterdiagram is present for the corresponding two-loop diagram. Note that $j_V, j_A$ do not mix, so the vector and axial-vector projections come from the vector and axial-vector insertions, respectively.
 
\begin{table}[H]
\centering
\renewcommand{\arraystretch}{1.1}
\resizebox{0.95\textwidth}{!}{%
\begin{tabular}{|l|c|c|c|c|c|c|c|c|}
\toprule
\textbf{Diagram} & \textbf{Color factor} & \textbf{Projection} & $A$ & $(B)_{_\text{NDR}}$ & $(B)_{_\text{BMHV}}$ & $\tilde{A}$   & $(\tilde{B})_{_\text{NDR}}$  & $(\tilde{B})_{_\text{BMHV}}$  \\
\midrule
J1 & * & * & $0$ & $0$  & $0$  & $0$  & $0$ & $0$ \\\hline
\multirow{2}{*}{J2[3]} & \multirow{2}{*}{$C_F^2$}  & $j_V$ &   $-1/2$ & $-5/4$  & $-5/4$  & $1$ & $1$ & $1$ \\
& & $j_A$ &   $-1/2$ & $-5/4$  & $-5/4$  & $1$ & $1$ & $5$ \\\hline
\multirow{2}{*}{J3[3]} & \multirow{2}{*}{$C_F^2 - C_A C_F/2$} & $j_V$ & $1$ & $3/2$ & $3/2$ & $-2$  & $-2$ & $-2$  \\
 &  & $j_A$ & $1$ & $3/2$ & $11/2$ & $-2$  & $-2$ & $-10$  \\\hline
\multirow{2}{*}{J4[2]} & \multirow{2}{*}{$C_AC_F$} & $j_V$ & $3/2$ & $23/4$  &  $23/4$ & $-3$ &  $-3$  & $-3$  \\
 &  & $j_A$ & $3/2$ & $23/4$  &  $47/4$ & $-3$ &  $-3$  & $-15$  \\\hline
  \multirow{2}{*}{J5} &  \multirow{2}{*}{$C_A C_F$}  & $j_V$ &   $-1/8$ & $13/16$  & $13/16$  &  & & \\
  &  & $j_A$ &   $-1/8$ & $13/16$  & $191/48$  &  & & \\\cline{1-6}
 \multirow{2}{*}{J6} &  \multirow{2}{*}{$C_A C_F$}  & $j_V$ &   $1/8$ & $7/16$  & $1/4$  &  & & \\
&   & $j_A$ &   $1/8$ & $7/16$  & $29/48$  &  & & \\\cline{1-6}
\multirow{2}{*}{J7} & \multirow{2}{*}{$C_FT_F$} & $j_V$ & $0$ & $-n_F$ & $-n_F$ &  & & \\
 &  & $ j_A$ & $0$ & $-n_F$ & $-11n_F/3$ &  & & \\ \hline
 \multirow{2}{*}{J8} & \multirow{2}{*}{$C_FT_F$}  & $j_V$ &   $-2$ & $-14/3$  & $-14/3$  & \multirow{2}{*}{-} & \multirow{2}{*}{-} & \multirow{2}{*}{-} \\
 &   & $j_A$ &   $0$ & $-3$  & $-3$  &  & & \\\hline
 \multirow{2}{*}{J9} & \multirow{2}{*}{$C_FT_F$}  & $j_V$ &   $2$ & $14/3$  & $14/3$  & \multirow{2}{*}{-} & \multirow{2}{*}{-} & \multirow{2}{*}{-} \\
 &   & $j_A$ &   $0$ & $-3$  & $-3$  &  & & \\
\bottomrule
\end{tabular}
}
\caption{Pole coefficients of two-loop $\mathcal{O}(\alpha_s^2)$ current insertion diagrams in the NDR and BMHV schemes. The columns and entries are to be interpreted as before. 
}\label{tab:J2loop}
\end{table}
 
\begin{figure}[H]
\centering
\begin{subfigure}{0.33\textwidth}
\centering
\begin{tikzpicture}
    \begin{feynman}
    \vertex (a)  ;
    \vertex [right=1cm of a] (b);
    \vertex [right=0.7cm of b] (c);
    \vertex [right=0.8cm of c] (d);
    \vertex [right=0.7cm of d] (e);
    \vertex [right=1cm of e] (f)  ;
    \diagram* {
            (a) -- [fermion, insertion = 1] (b) -- [fermion] (c) -- [fermion] (d) -- [fermion] (e) -- [fermion] (f), (d) -- [gluon, half right] (c), (e) -- [gluon, half right] (b)
        };
        \end{feynman}
\end{tikzpicture}
\caption{EG2[4]}
\end{subfigure}\hspace{0.4cm}
\begin{subfigure}{0.33\textwidth}
\centering
\begin{tikzpicture}
    \begin{feynman}
    \vertex (a)  ;
    \vertex [right=1cm of a] (b);
    \vertex [right=0.7cm of b] (c);
    \vertex [right=0.8cm of c] (d);
    \vertex [right=0.7cm of d] (e);
    \vertex [right=1cm of e] (f)  ;
    \diagram* {
            (a) -- [fermion] (b) -- [insertion = 0.5] (c) -- [fermion] (d) -- [fermion] (e) -- [fermion] (f), (d) -- [gluon, half right] (c), (e) -- [gluon, half right] (b)
        };
        \end{feynman}
\end{tikzpicture}

\caption{EK2[3]}
\end{subfigure}
\caption{Crosses represent evanescent operator insertion, where `EG' and `EK' imply gluonic and kinetic evanescent insertion.}
\label{fig:ev_insertion}
\end{figure}
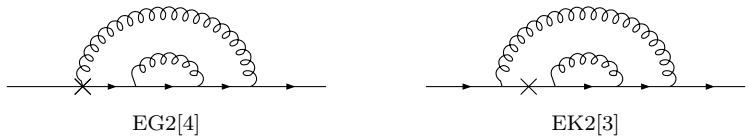

\begin{table}[H]
\centering
\renewcommand{\arraystretch}{1.1}
\resizebox{0.85\textwidth}{!}{%
\begin{tabular}{|l|c|c|c|c|c|c|c|}
\toprule
\multirow{2}{*}{\textbf{Diagram}} & \multirow{2}{*}{\textbf{Color factor}} & \multicolumn{2}{c|}{($\widetilde {\slashed p'} - \widetilde {\slashed p})\gamma^5$} & \multicolumn{4}{c|}{$(\widehat {\slashed p'} + \widehat{\slashed p})\gamma^5$} \\
\cline{3-8} 
& & $B$ & $\tilde B$ & $A$ & $B$ & $\tilde A$ & $\tilde B$ \\
\midrule
EG1[2], EK1 & * & $0$ & $0$  & $0$  & $0$  & $0$ & $0$ \\ \hline
EG2[4], EK2[3] & $C_F^2$  &    $0$ & $4$  & $-1/2$  & $-5/4$ & $1$ & $1$ \\ \hline
EG3[4], EK3[3] & $C_F^2 - C_AC_F/2$ &  $4$ & $-8$  &  $1$ & $3/2$ &  $-2$  & $-2$  \\
EG4[3], EK4[2] & $C_AC_F$ &  $6$ & $-12$ & $3/2$ & $23/4$ & $-3$ & $-3$  \\ \hline
EG5[2], EK5 & $C_A C_F$  &    $19/6$ & $-5/3$  & $-1/8$  & $13/16$ &\multirow{2}{*}{$-10/3$} & \multirow{2}{*}{$-20/3$} \\
EG6[2], EK6 & $C_A C_F$  &    $1/6$ & $14/3$  & $1/8$  & $7/16$ &  &  \\
EG7[4], EK7 &  $C_F T_F$  &    $-16n_F$ & $4n_F$  & $0$  & $-6n_F$ & $8n_F/3$ & $16n_F/3$ \\ \hline
EK8 & $C_FT_F$  &    $-3$ &  -  & $0$  & $0$ & - & - \\
EK9 & $C_FT_F$  &    $-3$ & - & $0$  & $0$ & - & - \\

\bottomrule
\end{tabular}
}
\caption{Pole coefficients of two-loop $\mathcal{O}(\alpha_s^2)$ diagrams with evanescent operator insertions in the BMHV scheme.}\label{tab:EV2loop}
\end{table}
The evanescent operator in the axial WI~\eqref{eq:bareWard} has terms with and without a gluon. The latter terms are inserted into the diagrams shown in Fig.~\ref{fig:2loopSEdia} at all possible places having a $\psi - G - \bar{\psi}$ vertex (Fig.~\ref{fig:ev_insertion} left panel). We denote such contributions by `EG$\#$' in table~\ref{tab:EV2loop}. The former purely kinetic-like evanescent terms are inserted in all diagrams shown in Fig.~\ref{fig:2loopJdia} in place of the current operator (Fig.~\ref{fig:ev_insertion} right panel). These are denoted by `EK$\#$'. The full evanescent vertex function is the sum of these two contributions. The pole coefficients for projections to the physical and evanescent operator structures are shown in table~\ref{tab:EV2loop}. In a particular row, the pole coefficient entries shown are the sum total of the specific EG and EK diagrams written in the leftmost column.

\section{Axial WI in NDR Scheme}\label{app:NDR}
In Naive Dimensional Regularization (NDR), an anticommuting $\gamma_5$ is chosen in $d$-dimensions~\cite{Chanowitz:1979zu} $\{\gamma^\mu, \gamma_5 \} = 0$. If we further assume the cyclicity of $d$-dimensional traces involving an odd number of $\gamma_5$, then NDR is unable to recover the correct result for $\text{tr}(\gamma^\mu \gamma^\nu \gamma^\rho \gamma^\sigma \gamma_5)$~\cite{Martin:1999cc}. To make the scheme practical, one formally demands the usual `trace' identity
\begin{equation}\label{eq:4gtrace}
    \text{tr}(\gamma^\mu \gamma^\nu \gamma^\rho \gamma^\sigma \gamma_5) = -4i\epsilon^{\mu\nu\rho\sigma}
\end{equation}
in $d$-dimensions~\cite{Isidori:2023pyp}. Anticommutativity, Eq.~\eqref{eq:4gtrace}, and trace-cyclicity are impossible to satisfy simultaneously in $d$-dimensions. In NDR, one keeps the former two at the expense of the latter. A $d$-dimensional `trace' operation is defined as a linear functional acting on Dirac matrices that in the limit $d\to 4$ becomes the usual trace operation but is non-cyclic in $d\neq 4$~\cite{Korner:1991sx}. This means that traces appearing in diagrams must be written with a consistent reading-point prescription. In our NDR calculations, we follow the original reading prescription of reading the anomaly traces $AVV$ (J8, J9 of Fig.~\ref{fig:2loopJdia}) starting from the axial-vertex. Once written in this manner, a simple way to compute the traces is to anticommute $\gamma_5$ to the rightmost and replace~\cite{Kreimer:1993bh, Ellis:2024omd}
\begin{equation}
    \gamma_5 \to -\frac{i}{24} \epsilon_{\mu\nu\rho\sigma} \gamma^\mu \gamma^\nu  \gamma^\rho \gamma^\sigma
\end{equation}
The resulting traces contain only $\gamma^\mu$ matrices and can be easily evaluated in NDR. The contraction with the Levi-Civita tensor is taken at the end of the calculations.

Ref.~\cite{Korner:1991sx} showed how the one-loop ABJ anomaly can be obtained in NDR using such a reading point prescription. Thereafter, it is widely believed that the four-dimensional axial Ward identity may hold in NDR to all orders without any extra finite renormalization. We explicitly check this by calculating matrix elements of $\overline{\text{MS}}$ renormalized operators on both sides of the Ward identity. Considering an external pair of fermions, the one-loop verification is straightforward. We evaluate the $\mathcal{O}(\alpha_s)$ current insertion diagram and self-energy contributions of Fig.~\ref{fig:oneloopdia}. The tree-level $G\tilde{G}$ matrix element between fermions is zero. At one-loop, this means that the renormalized Ward identity indeed holds without the need for a finite renormalization of the axial current. However, at two-loops $\mathcal{O}(\alpha_s^2)$, our explicit computations show an inequality between the left- and right-hand sides of the $\overline{\text{MS}}$-renormalized operator matrix elements. In this case, the one-loop matrix element of $G\tilde{G}$ between external fermions also enters the computation. The self-energy diagrams are given in Fig.~\ref{fig:2loopSEdia} and the current insertion diagrams in Fig.~\ref{fig:2loopJdia}. We refer the reader to~\cite{Chen:2023lus} for a thorough discussion of this point, where such a violation of the axial Ward identity was noted for the first time using matrix elements with gluon external states.

\bibliographystyle{JHEP}
\bibliography{referencesHV}

\end{document}